\documentstyle[12pt,amsmath,amssymb,epsf,epic]{article}

\numberwithin{equation}{section}

\newcommand{\un}{{\mathbb I}}
\newcommand{\ra}{\rightarrow}
\newcommand{\tr}{\mbox{Tr}}

\newcommand{\spr}{\mbox{Spr}\,} 
\newcommand{\bra}{\langle} 
\newcommand{\ket}{\rangle}

\newcommand{\E}{{\mathbb E}}
\newcommand{\D}{{\mathbb D}}
\newcommand{\M}{{\mathbb M}}

\newcommand{\be}{\begin{equation}}
\newcommand{\ee}{\end{equation}}
\newcommand{\bea}{\begin{eqnarray}}
\newcommand{\eea}{\end{eqnarray}}

\newcommand{\eps}{\epsilon}

\newcommand{\ffi}{\varphi}

\newcommand{\ep}{\hfill  {\vrule height 10pt width 8pt depth 0pt}}

\newcommand{\grintl}{[\kern-.18em [}
\newcommand{\grintr}{]\kern-.18em ]}

\newcounter{resultcounter}[section]

\newtheorem{thm}[resultcounter]{Theorem}
\newtheorem{lem}[resultcounter]{Lemma}
\newtheorem{prop}[resultcounter]{Proposition}
\newtheorem{cor}[resultcounter]{Corollary}
\newtheorem{definition}[resultcounter]{Definition}
\newtheorem{rem}[resultcounter]{Remark}

\def\bed{\begin{definition}}
\def\eed{\end{definition}}

\def\proof{\noindent{\bf Proof.}\ \ }

 \def\cB{{\cal B}} 
 \def\cE{{\cal E}} \def\cF{{\cal F}}
 \def\cH{{\cal H}} \def\cI{{\cal I}}
\def\cJ{{\cal J}}  
\def\cM{{\cal M}} \def\cN{{\cal N}} 
  
\def\cS{{\cal S}} \def\cT{{\cal T}}

\newcommand{\R}{{\mathbb R}}
\newcommand{\N}{{\mathbb N}}

\newcommand{\C}{{\mathbb C}}
\newcommand{\Z}{{\mathbb Z}}
\renewcommand{\E}{{\mathbb E}}
\renewcommand{\P}{{\mathbb P}}
\newcommand{\I}{{\mathbb I}}
\newcommand{\T}{{\mathbb T}}

\def\proof{\noindent{\bf Proof:}\ \ }

\begin{document}
\title{
Random Time-Dependent Quantum Walks}
 \author{ Alain Joye\footnote{ Institut Fourier, UMR 5582,
CNRS-Universit\'e Grenoble I, BP 74, 38402 Saint-Martin
d'H\`eres, France.} \footnote{Partially supported by the Agence Nationale de la Recherche, grant ANR-09-BLAN-0098-01}}

\date{ }

\maketitle
\vspace{-1cm}
\abstract{
We consider the discrete time unitary dynamics given by a quantum walk on the lattice $\Z^d$ performed by a quantum particle with internal degree of freedom, called coin state,  according to the following iterated rule: a unitary update of the coin state takes place, followed by a shift on the lattice, conditioned on the coin state of the particle. We study the large time behavior of the quantum mechanical probability distribution of the position observable in $\Z^d$ when the sequence of unitary updates is given by an i.i.d. sequence of random matrices. When averaged over the randomness, this distribution is shown to display a drift proportional to the time and its centered counterpart is shown to display a diffusive behavior with a diffusion matrix we compute. A moderate deviation principle is also proven to hold for the averaged distribution and the limit of the suitably rescaled corresponding characteristic function is shown to satisfy a diffusion equation. A generalization to unitary updates distributed according to a Markov process is also provided.

An example of i.i.d. random updates for which the analysis of the distribution can be performed without averaging is worked out. The distribution also displays a deterministic drift proportional to time and its centered counterpart gives rise to a random diffusion matrix whose law we compute. A large deviation principle is shown to hold for this example. We finally show that, in general, the expectation of the random diffusion matrix equals the diffusion matrix of the averaged distribution.

}

\thispagestyle{empty}
\setcounter{page}{1}
\setcounter{section}{1}

\setcounter{section}{0}

\section{Introduction}

Quantum walks are models of discrete time quantum evolution taking place on a $d$-dimensional lattice. Their implementation as unitary discrete dynamical systems on a Hilbert space is typically the following. A quantum particle with internal degree of freedom moves on an infinite $d$-dimensional lattice according to the following rule. The one-step motion consists in an update of the internal degree of freedom by means of a unitary transform in the relevant part of the Hilbert space followed by a finite range shift on the lattice, conditioned on the internal degree of freedom of the particle. Quantum walks constructed this way can be considered as quantum analogs of classical random walks on lattices. Therefore, in this context, the space of the internal degree of freedom is called {\it coin space}, the degree of freedom is the {\it coin state} and the unitary operators performing the update are {\it coin matrices}.

Due to the important role played by classical random walks in theoretical computer science, quantum walks have enjoyed an increasing popularity in the quantum computing community in the recent years, see for example \cite{M},  \cite{AAKV},  \cite{Ke}, \cite{Ko}. Their particular features for search algorithm is described in \cite{SKW}, \cite{AKR}, \cite{MNRS} and in the review \cite{S}. 
In addition, quantum walks can be considered as effective dynamics of quantum systems in certain asymptotic regimes. See {\em e. g.} \cite{cc}, \cite{ADZ}, \cite{M}, \cite{lv},  \cite{bb}, \cite{rhk}, for a few models of this type, and \cite{ade}, \cite{BHJ}, \cite{dOS}, \cite{HJS}, \cite{abj} for their mathematical analysis.
Moreover, quantum walk dynamics have been shown to be an experimental reality for systems of cold atoms trapped in suitably monitored optical lattices  \cite{Ketal}, and ions caught in monitored Paul traps \cite{Zetal}.

\medskip

While several variants and generalizations of the quantum dynamics described above are possible, we will focus on the case where the underlying  lattice is $\Z^d$ and where the dimension of the coin space is $2d$. We are interested in the long time behavior of quantum mechanical expectation values of observables that are non-trivial on the lattice only, {\em i.e.} that do not depend on the internal degree of freedom of the quantum walker. Equivalently, this amounts to studying a family of random vectors $X_n$ on the lattice $\Z^d$, indexed by the discrete time variable, with probability laws $\P(X_n=k)=W_k(n)$ defined by the prescriptions of quantum mechanics. The initial state of the quantum walker is described by a density matrix.

The case where the unitary update of the coin variable is performed at each time step by means of the same coin matrix is well known. It leads to a ballistic behavior of the expectation of the position variable characterized by $\E_{W(n)}(X_n)\simeq n V$ when $n$ is large, for some vector $V$. This vector and further properties of the motion can be read off the Fourier transform of the one step unitary evolution operator.

\bigskip

In this paper, we consider the situation where the coin matrices used to update the coin variable depend on the time step in a random fashion, that is a situation of {\it temporal disorder}. Let us describe our results informally here, referring the reader to the relevant sections for precise statements.

We assume the sequence of coin matrices consists of random unitary matrices which are independent and identically distributed (i.i.d.) and we analyze the large $n$ behavior of the corresponding random distribution $W^\omega(n)$ of $X_n^\omega$. We do so by studying the characteristic function $\Phi_n^\omega(y)=\E_{W^\omega(n)} (e^{iyX_n^\omega})$. 
In Section \ref{dets}, we first show a deterministic result saying that the characteristic function at time $n$ can be expressed in terms of a product of $n$ matrices, $M_j$, each $M_j$ depending on the coin operator at step $j$ only,  in the spirit of the GNS construction, see Propositions \ref{phiphi}, \ref{phiden}. In the random case, the $M_j$'s become i.i.d. random matrices $M_{\omega}$.

Then we address the behavior of the {\it averaged} distribution $w(n)=\E_\omega(W^\omega(n))$ of $X_n$, for $n$ large in Section \ref{rands}. Theorem \ref{cf} says  under certain natural spectral assumptions on the matrices $E_\omega(M_\omega)$ that  $X_n$ displays a ballistic behavior 
$$\E_{w(n)}(X_n)\simeq n \overline r \in \R^d$$ 
where $\overline r$ is a drift vector depending only on the properties of the deterministic shift operation following the random update of the coin state. Moreover, the centered random vector $(X_n-n\overline r)$ is shown to display a diffusive behavior characterized by a diffusion matrix $\D$ we compute: 
$$
\E_{w(n)}((X_n-n\overline r)_i(X_n-n\overline r)_j)\simeq n\D_{ij}, \ i,j =1,2, \cdots, d.
$$
We also show in Theorem \ref{cf} that for any $t>0$, $y\in \R^d$, the averaged and rescaled characteristic function 
$e^{-i[tn]\overline ry/\sqrt{n}}\E_\omega(\Phi_{[tn]}^\omega(y/\sqrt{n}))$ converges for large $n$, in a certain sense, to the Fourier transform of superpositions of solutions to a diffusion equation, with diffusion matrix $\D(v)$, $v\in \T^d$, the $d$-dimensional torus:
$$
e^{-i[tn]\overline ry/\sqrt{n}}\E_\omega(\Phi_{[tn]}^\omega(y/\sqrt{n}))\ra \int_{\T^d}e^{-\frac t2\bra y | \D(v) y\ket}{dv}/{(2\pi)^d}.
$$

In Section \ref{ein}, we briefly discuss the relationship between the drift vector $\overline r$ and the diffusion matrix $\D$ in case the deterministic shift can take arbitrarily large values.
Then we investigate finer properties of the behavior in $n$ of the averaged distribution in Section \ref{moder}. Theorem \ref{md} states that a {\it moderate deviation principle} holds for $w(n)$: there exists a {\it rate function} $\Lambda^*:\R^d\ra[0,\infty]$ such that, for any set $\Gamma\in \R^d$ and any  $0<\alpha<1$, as $n\ra\infty$, 
\be
\P(X_n-n\overline r\in n^{(\alpha +1)/2}\, \Gamma)\simeq e^{-n^\alpha \inf_{x\in {\Gamma}}\Lambda^*(x)}.
\ee

In Section \ref{four}, we consider a distribution of coin operators which allows us to analyze the random distribution $W^\omega(n)$, without averaging over the temporal disorder. This distribution is supported, essentially, on the unitary permutation matrices. We show that in this case $W^\omega(n)$ coincide with  the distribution of a Markov chain with finite state space whose transition matrix we compute explicitly.    Consequently, we get that the centered random vector $X_n^\omega-n\overline r$ converges in distribution to a normal law $\cN(0,\Sigma)$, with an explicit  correlation matrix $\Sigma$, and an explicit deterministic drift vector $\overline r$ given in Theorem \ref{clt}. In turn, this allows us to show in  Corollary \ref{randd} the existence of a random diffusion matrix $\D^\omega$ such that
$$
\E_{W^\omega(n)}((X_n^\omega-n\overline r)_i(X_n^\omega-n\overline r)_j)\simeq n\D^\omega_{ij}, \ i,j =1,2, \cdots, d,
$$
whose matrix elements $\D^\omega_{ij}$ are distributed according to the law of $X_i^\omega X_j^\omega$, where the vector $X^\omega$ is distributed according to $\cN(0,\Sigma)$. Finally, a large deviation principle for the random distribution $W^\omega(n)$ is stated as Theorem \ref{ldp}. This example also shows that we cannot expect almost sure convergence results for random quantum walks. 
\medskip 

We close the paper by showing how to generalize the results of Sections \ref{rands} and \ref{moder}  to the case where the random coin matrices are not independent anymore and are distributed according to a Markov process with a finite number of states. See Section \ref{mark}.

\bigskip

Let us comment about the literature. In a sense, the situation we address corresponds to the cases considered in \cite{p}, \cite{ks}, \cite{hks} where the dynamics is generated by a quantum Hamiltonian with a time dependent potential generated by a random process. For quantum walks, the role of the random time dependent potential is played by the random coin operators whereas the role of the deterministic kinetic energy is played by the shift. 

Quantum walks with unitary random coin operators have been tackled in some numerical works, see \cite{SBBH}, for example. On the analytical side, we can mention \cite{KBH} (see also \cite{jm}) where particular hypotheses on the coin matrices reduce the problem to the study of correlated random walks. During the completion of the paper, the preprint \cite{avww} appeared. It reviews and addresses several types of quantum walks, deterministic and random, decoherent and unitary. In particular, the averaged dynamics of random quantum walks of the type studied in the present paper are tackled, by means of a similar approach. The results we prove, however, are more detailed and go beyond those of \cite{avww}.

We finally note that there exist another instance of random quantum walks in which the randomness lies in space rather than in time. In the present context, this means the coin operators depend on the sites of the lattice and are chosen according to some law, in the same spirit as for the Anderson model. Dynamical or spectral localization are the phenomena on interest there. See for example \cite{Ko1},\cite{jm}, \cite{sk} and references therein for results about such questions.

\medskip

\medskip

{\bf Acknowledgements} It is a pleasure to thank L. Bruneau, E. Hamza, M. Merkli and C.A. Pillet for fruitful discussions and suggestions about  this work.

\section{General Setup}\label{dets}

Let $\cH=\C^{2d}\otimes l^2(\Z^{d})$ be the Hilbert space of the quantum walker in $\Z^d$ with $2d$ internal degrees of freedom. We denote the canonical basis of $\C^{2d}$ by $\{|\tau\ket \}_{\tau\in I^d_\pm}$, where $I_\pm=\{\pm 1, \pm 2, \dots, \pm d\}$, so that the orthogonal projectors on the basis vectors are noted  $P_\tau=|\tau\ket\bra \tau|$, $\tau \in I_\pm$.  We 
denote the canonical basis of $l^2(\Z^{d})$ by $\{|x\ket\}_{x\in\Z^d}$. We shall write for a vector $\psi\in \cH$,  $\psi=\sum_{x\in\Z^d}\psi(x)|x\ket$, where $\psi(x)=\bra x|\psi\ket\in \C^{2d}$ and $\sum_{x\in\Z^d}\|\psi(x)\|_{\C^{2d}}^2=\|\psi\|^2<\infty$. We shall abuse notations by using the same symbols $\bra \cdot | \cdot\ket$ for scalar products and corresponding "bra" and "ket" vectors on $\cH$, $\C^{2d}$ and $l^2(\Z^d)$, the context allowing us to determine which spaces we are talking about. Also, we will  often drop the subscript ${\C^{2d}}$ of the norm.

\medskip

A {\it coin matrix} acting on the internal degrees of freedom, or {\it coin state}, is a unitary matrix $C\in M_{2d}(\C)$ and a {\it jump function} is a function $r : I_\pm\ra\Z^d$.

The corresponding one step unitary evolution $U$  of the walker on $\cH=\C^{2d}\otimes l^2(\Z^{d})$ is given by 
\be
U=S\ (C\otimes \I),
\ee
where $\I$ denotes the identity operator and the shift $S$ is defined on $\cH$ by
\bea\nonumber
S&=&\sum_{x\in \Z^d}\sum_{\tau\in \{1,\cdots, d\}} P_\tau\otimes |x+r(\tau)\ket\bra x|+P_{-\tau}\otimes |x+r(-\tau)\ket\bra x|\\
&=&\sum_{x\in \Z^d}\sum_{\tau\in I_\pm} P_\tau\otimes |x+r(\tau)\ket\bra x|.
\eea
By construction, a walker at site $y$ with internal degree of freedom $\tau$ represented by the vector $|\tau\ket\otimes |y\ket\in \cH$ is just sent by $S$ to one of the neighboring sites depending on $\tau$ determined by the jump function $r(\tau)$
\be
S\ |\tau\ket\otimes |y\ket=|\tau\ket\otimes |y+r(\tau)\ket. 
\ee
The composition by $C\otimes \I$ reshuffles or updates the coin state so that the pieces of the wave function corresponding to different internal states are shifted to different directions, depending on the internal state. We can write 
\be
U=\sum_{x\in \Z^d}\sum_{\tau\in I_\pm} P_\tau C\otimes |x+r(\tau)\ket\bra x|.
\ee

\medskip

Given a set of $n>0$ unitary coin matrices $C_k\in M_{2d}(\C)$, $k=1,\cdots, n$, we define the corresponding discrete evolution from time zero to time $n$ by
\be\label{evoln}
U(n,0)=U_n U_{n-1}\cdots U_1, \ \ \mbox{where} \ \ U_k=S\ (C_k\otimes \I).
\ee

\medskip

Let $f:\Z^d\ra \C$ and define the multiplication operator $F:D(F)\ra \cH$ on its domain $D(F)\subset\cH$ by $(F\psi)(x)=f(x)\psi(x)$, $\forall  x\in\Z^d$, where $\psi\in D(F)$ is equivalent to $\sum_{x\in \Z^d}|f(x)|^2\|\psi(x)\|_{\C^d}^2<\infty$. Note that $F$ acts trivially on the coin state. 

When $f$ is real valued, $F$ is self-adjoint and will be called a {\it lattice observable}.

\subsection{Vector states}

In particular, consider a walker characterized at time zero by the normalized vector $\psi_0=\ffi_0\otimes |0\ket$, {\em i.e.} which sits on site $0$ with coin state $\ffi_0$. The quantum mechanical expectation value of a lattice observable $F$ at time $n$ is given by $\bra F \ket_{\psi_0}(n)=\bra \psi_0|U(n,0)^* F U(n,0)\psi_0\ket$.

As in \cite{jm}, a straightforward computation yields 
\begin{lem}\label{expn} With the notations above, 
\bea\nonumber
U(n,0)&=&\sum_{x\in \Z^d}\sum_{\tau_1, \tau_2, \dots, \tau_n\in {I_\pm}^n} P_{\tau_n}C_{n}P_{\tau_{n-1}}C_{n-1}\cdots P_{\tau_1}C_{1} \\ \nonumber
& & \quad \quad \quad  \otimes |x+r(\tau_1)+\cdots +r(\tau_n)\ket\bra x | \\
&\equiv&\sum_{x\in \Z^d}\sum_{k\in \Z^d}J_k(n)\otimes |x+k\ket\bra x|,
\eea
where 
\be
J_k(n)=\sum_{\tau_1, \tau_2, \dots, \tau_n\in {I_\pm}^n\atop \sum_{s=1}^nr(\tau_s)=k} P_{\tau_n}C_{n}P_{\tau_{n-1}}C_{n-1}\cdots P_{\tau_1}C_{1}\in M_{2d}(\C)
\ee
and $J_k(n)=0$, if $\sum_{s=1}^nr(\tau_s)\neq k$.
Moreover, for any lattice observable $F$, and any normalized vector $\psi_0=\ffi_0\otimes |0\ket$, 
\bea\label{rvlat}\nonumber
\bra F \ket_{\psi_0}(n)&=&\bra \psi_0|U^*(n,0)F U(n,0)\psi_0\ket=\sum_{k\in\Z^d}f(k)\bra\ffi_0|J_k(n)^*J_k(n)\ffi_0\ket\\
&\equiv&\sum_{k\in\Z^d}f(k)W_k(n),
\eea
where $W_k(n)=\|J_k(n)\ffi_0\|^2_{\C^{2d}}$ satisfy
\be\label{norm}
\sum_{k\in\Z^d}W_k(n)=\sum_{k\in\Z^d}\|J_k(n)\ffi_0\|^2_{\C^{2d}}=\|\psi_0\|^2_{\cH}=1.
\ee
\end{lem}
\begin{rem}\label{defx} We view the non-negative quantities $\{W_k(n)\}_{n\in\N^*}$ as the probability distributions of a sequence of $\Z^d$-valued random variables $\{X_n\}_{n\in\N^*}$, with 
\be
\mbox{\em Prob}(X_n=k)=W_k(n)= \bra\psi_0 | U(n,0)^* (\I\otimes |k\ket\bra k|) U(n,0)\psi_0\ket=\|J_k(n)\ffi_0\|^2_{\C^{2d}},
\ee
in keeping with (\ref{rvlat}). In particular,  $\bra F \ket_{\psi_0}(n)=\E_{W_k(n)}(f(X_n))$. We shall use freely both notations.

\end{rem}
\begin{rem}
\medskip 
All sums over $k\in\Z^k$ are finite since $J_k(n)=0$ if $\max_{j=1,\dots,d}|k_j|>\rho n$, for some $\rho>0$ since the jump functions have finite range.

\end{rem}
We are particularly interested in the long time behavior, $n >\hspace{-4pt}> 1$,  of $\bra X^2\ket_{\psi_0}(n)$, the expectation of the observable $X^2$ corresponding to the function $f(x)=x^2$ on $\Z^d$ with initial condition $\psi_0$. Or, in other words, in the second moments of the distributions $\{W_k(n)\}_{n\in\N^*}$.

\medskip

Let us proceed by expressing the probabilities $W_k(n)$ in terms of the $C_k$'s, $k=1,\dots, n$. We need to introduce some more notations. Let $I_n(k)=\{\tau_1,\cdots, \tau_n\}$, where $\tau_l\in I_\pm$, $l=1,\dots,n$ and $\sum_{l=1}^n r(\tau_l)=k$. In other words, $I_n(k)$ denotes the set of paths that link the origin to $k\in \Z^d$ in $n$ steps via the jump function $r$. Let us write $\ffi_0=\sum_{\tau\in I_\pm}a_\tau |\tau\ket$.
\begin{lem}\label{wkn}
\be
W_k(n)=\sum_{{\tau_0,{\{\tau_1,\cdots, \tau_n\}\in I_n(k)} \atop {\tau_0', \{\tau_1',\cdots, \tau_n'\}\in I_n(k)}} \atop \mbox{\tiny s.t.}\ \tau_n=\tau_n' }\overline{a_{\tau_0'}}a_{\tau_0}  \bra\tau_0'|C_1^*\ \tau_1'\ket\bra \tau_1|C_1\tau_0\ket \prod_{s=2}^n \bra \tau_{s-1}'|C_s^*\ \tau_s'\ket\bra \tau_s|C_s \ \tau_{s-1}\ket .
\ee
\end{lem}

We approach the problem through the characteristic functions $\Phi_n$ of the probability distributions $\{W_\cdot(n)\}_{n\in\N^*}$ defined by the periodic function
\be\label{char}
\Phi_n(y)=\E_{W(n)}(e^{i yX_n})=\sum_{k\in \Z^d}W_k(n)e^{i yk}, \ \ \mbox{where }\ y\in [0,2\pi)^d.
\ee
To emphasize the dependence in the initial state, we will sometimes write $\Phi_n^{\ffi_0}$ and/or $W_k^{\ffi_0}(n)$.
All periodic functions will be viewed as  functions defined on the torus, {\it i.e.} $[0,2\pi)^d \simeq \T^d.$
The asymptotic properties of the quantum walk emerge from the analysis of the limit in an appropriate sense as $n\ra\infty$ of the characteristic function in the {\em diffusive scaling}
\be\label{diffscal}
\lim_{n\ra\infty}\Phi_n(y/\sqrt n)
\ee

Looking for a relation between $W_k(n)$ and $W_k(n+1)$, we find that the condition $\tau_n=\tau_n'$ is a nuisance we can relax now and deal with later. 

\medskip

Consider the set of paths $G_n(K)$ in $\Z^{2d}$ from the origin to $K=\begin{pmatrix}k \\ k' \end{pmatrix}\in \Z^{2d}$ via the (extended) jump function defined by  
\be R:I_\pm^2\ra Z^{2d}, \ \ \ R\begin{pmatrix}\tau_s \\ \tau_s' \end{pmatrix}=\begin{pmatrix}r(\tau_s) \\ r(\tau_s') \end{pmatrix},
\ee that is paths of the form $(T_1, \cdots,  T_{n-1}, T_n)$, where $T_s=\begin{pmatrix}\tau_s \\ \tau_s' \end{pmatrix}\in I_\pm^2$, $s=1,2, \dots,n$, and $ \sum_{s=1}^n R(T_s)=K$.
Then note that the generic term in Lemma \ref{wkn} reads 
\be
\bra \tau_{s-1}'|C_s^*\ \tau_s'\ket\bra \tau_s|C_s \ \tau_{s-1}\ket = \overline{\bra \tau_{s}'|C_s\ \tau_{s-1}'\ket}\bra \tau_s|C_s \ \tau_{s-1}\ket\equiv\bra \tau_s\otimes\tau'_{s} | (C_s\otimes \overline C_s) \ \tau_{s-1}\otimes\tau'_{s-1} \ket,
\ee
where, in the last expression, we introduced the unitary tensor product   
\be\label{tpv}
V(s)\equiv C_s\otimes \overline C_s \ \ \mbox{in}\ \C^{2d}\otimes \C^{2d}.
\ee 
The canonical basis of $\C^{2d}\otimes \C^{2d}$ is $\{ |\tau\otimes\tau'\ket\}_{(\tau, \tau')\in I_\pm^2}$, hence, with the identification 
\be
T=\begin{pmatrix}\tau \\ \tau' \end{pmatrix}\simeq |\tau\otimes\tau'\ket ,
\ee 
we can write the matrix elements of $V(s)$ as
\be
\bra \sigma \otimes\sigma' | (C_s\otimes \overline C_s) \ \tau\otimes\tau' \ket\equiv V(s)_{S T}, \ \ \ S=\begin{pmatrix}\sigma \\ \sigma' \end{pmatrix}
, T=\begin{pmatrix}\tau \\ \tau' \end{pmatrix}\in I^2_\pm.
\ee
With the decompositions
\be
\ffi_0=\sum_{\tau\in I_\pm}a_\tau |\tau\ket \ \Rightarrow \ \chi_0=\ffi_0\otimes\overline{\ffi_0}=\sum_{(\tau, \tau')\in I^2_\pm}a_\tau \overline a_{\tau'}|\tau\otimes \tau'\ket 
\ee
we can write
\be\label{init}
\bra T_1| V(1){\chi_0}\ket=\sum_{T_0 \in I^2_\pm}V(1)_{T_1T_0}A_{T_0}, \ \ \mbox{where} \ A_{T_0}=a_{\tau_0}\overline a_{\tau'_0}, \ \mbox{for} \ T_0=\begin{pmatrix}\tau_0 \\ \tau'_0 \end{pmatrix}.
\ee

With these notations, we consider the {\it weight} of $n$-step paths in $\Z^{2d}$ from the origin to $K$, with last step $T$, defined by
\be
W_K^T(n)=\sum_{(T_1, \cdots,  T_{n-1})\in {I^2_\pm}^{n-1} \ \mbox{\tiny s.t.} \atop {(T_1, \cdots,  T_{n-1}, T)\in G_n(K)}}V(n)_{T T_{n-1}}\cdots V(2)_{T_2 T_1}V(1)_{T_1\chi_0}.
\ee

Note that by construction, see Lemma \ref{wkn},
\be\label{wkn2}
W_k(n)=\sum_{T\in H_\pm}W^T_{K_k}(n)\ \ \mbox{with } \ K_k=\begin{pmatrix} k \\ k \end{pmatrix} \ \mbox{and } \ H_\pm=\left\{\begin{pmatrix} \tau \\ \tau \end{pmatrix}, \tau\in I_\pm \right\}. 
\ee
We also introduce corresponding periodic functions $\Phi_n^T$, $n>0$ by
\be\label{gphy}
\Phi_n^T(Y)=\sum_{K\in\Z^{2d}}e^{iYK}W_K^T(n), \ \ \mbox{where} \ Y\in \T^{2d}
\ee
and, see (\ref{init}), 
\be
\Phi_0^T(Y)=A_T.
\ee

These definitions lead to the sought for relationships:
\begin{prop}
For all $n\in \N$, $T\in I^2_\pm$, $K\in \Z^{2d}$ and $Y\in \T^{2d} $,
\bea
W_K^T(n+1)&=&\sum_{S\in I^2_\pm} V(n+1)_{TS}W_{K-R(T)}^S(n),\\
\Phi_{n+1}^T(Y)&=&\sum_{S\in I^2_\pm}e^{iYR(T)}V(n+1)_{TS}\Phi_{n}^S(Y).
\eea
\end{prop}

We can express these relationships in a yet more concise way as follows. Recall that for each basis vector $T\simeq \tau\otimes\tau'$ and each vector $Y=(y,y')\in \T^{d}\times  \T^{d}$, we have
\be\label{convention}
YR(T)=yr(\tau)+y'r(\tau')\in \R. 
\ee
Introduce the vectors in $\C^{4d^2}\simeq \C^{2d}\otimes \C^{2d}$ with $Y\in \T^{2d} $ and $n\geq 0$
\be
 {\bf \Phi}_n(Y) = \sum_{T=(\tau,\tau')\in I_\pm^2} \Phi_n^{T}(Y)\, |\tau\otimes\tau' \ket \ \ \mbox{and }\ {\bf \Phi}_0 =  \sum_{T=(\tau,\tau')\in I_\pm^2} A_T\, |\tau\otimes\tau' \ket
\ee
where ${\bf \Phi}_0$ is determined by the internal state $\ffi_0$ only and is independent of $Y$. 

With the matrices on  $\C^{2d}\otimes \C^{2d}\simeq \C^{4d^2}$ expressed in the ordered basis  $I^2_\pm$ by
\be
D(Y)=\sum_{T=(\tau,\tau')\in I_\pm^2} e^{iYR(T)} \, |\tau\otimes\tau' \ket \bra \tau\otimes\tau' |,  \ \ \mbox{with }\ Y\in \T^{2d}, 
\ee 
and
\be
V(s)=C_s\otimes \overline C_s, \ \ M_s(Y)=D(Y)V(s),
\ee
we get the 
\begin{cor}\label{prodM}For any $n\geq 0$ and $Y\in \T^{2d}$,
\bea
{\bf \Phi}_n(Y)&=&M_n(Y)M_{n-1}(Y)\cdots M_1(Y){\bf \Phi}_0.
\eea
\end{cor}
\begin{rem} The matrix $D(Y)$ can be expressed as a tensor product of unitary diagonal matrices. Let $Y=(y,y')\in  \T^d\times  \T^d$. Then
\bea\label{defd}
D(y,y')&=&\sum_{(\tau,\tau')\in I^2_\pm}e^{i(yr(\tau)+y'r(\tau'))}|\tau\otimes\tau'\ket\bra\tau\otimes\tau'|=\sum_{\tau\in I_\pm}e^{iyr(\tau)}|\tau\ket\bra\tau|\otimes \sum_{\tau'\in I_\pm}e^{iy'r(\tau')}|\tau'\ket\bra\tau'|\nonumber\\
&\equiv& d(y)\otimes d(y')\ \ \mbox{where }\ d(y)=\sum_{\tau\in I_\pm}e^{iyr(\tau)}|\tau\ket\bra\tau|.
\eea
Consequently, we can write
\be\label{tpm}
M_s(y,y')=d(y)C_s\otimes \overline{d(-y')C_s}.
\ee
Together with the fact that ${\bf \Phi}_0=\ffi_0\otimes\overline\ffi_0$, this yields
\bea\label{tsfj}
{\bf \Phi}_n(y,y')&=&d(y)C_n \cdots d(y)C_1\, \ffi_0\otimes \overline{d(-y')C_nd(-y') \cdots C_1\, \ffi_0}\nonumber\\
&\equiv&\cJ_n(y)\ffi_0\otimes\overline{\cJ_n(-y')\ffi_0} =(\cJ_n(y)\otimes \overline{\cJ_n(-y')})\, \ffi_0\otimes\overline{\ffi_0}.
\eea
\end{rem}

We note here for future reference that $\cJ_n(y)$ is the Fourier transform of $J_k(n)$:
\begin{lem}\label{Jfourier}
For any $y\in\T^d$ 
\be
\cJ_n(y)=\sum_{k\in\Z^d}e^{iyk}J_k(n).
\ee
\end{lem}
\proof With the convention (\ref{convention}), the right hand side reads
\bea
&&\sum_{k\in\Z^d} \sum_{(\tau_1, \tau_2, \dots, \tau_n)\in {I_\pm}^n\atop \sum_{s=1}^nr(\tau_s)=k} e^{iy(r(\tau_1)+\cdots+r(\tau_n))}P_{\tau_n}C_{n}P_{\tau_{n-1}}C_{n-1}\cdots P_{\tau_1}C_{1}\nonumber\\
&&= \sum_{(\tau_1, \tau_2, \dots, \tau_n)\in {I_\pm}^n}e^{iyr(\tau_n)}|\tau_n\ket\bra\tau_n|C_n e^{iyr(\tau_{n-1})}|\tau_{n-1}\ket\bra\tau_{n-1}|C_{n-1} \cdots e^{iyr(\tau_1)}|\tau_1\ket\bra\tau_1|C_1\nonumber\\
&&=d(y)C_n d(y) C_{n-1}\cdots d(y)C_1=\cJ_n(y).
\eea
\ep

\medskip

Eventually, the characteristic function $\Phi_n(y)$ we are interested in, see (\ref{char}), can be obtained from ${\bf \Phi}_n(Y)$. We shall denote the normalized measure on the torus $\T^d$ by $d{\tilde v}=\frac{dv}{(2\pi)^d}$.

\begin{prop}\label{phiphi}
For any $n\in \N$ and $y\in \T^d$, 
\be
\Phi^{\ffi_0}_n(y)
=\int_{\T^d}\bra{\bf \Psi}_1 |M_n(y-v,v)M_{n-1}(y-v,v)\cdots M_1(y-v,v){\bf \Phi}_0\ket \, d{\tilde v}, 
\ee where
\be
{\bf \Psi}_1=\sum_{T\in H_\pm} |T\ket=\sum_{\tau\in I_\pm}|\tau\otimes\tau\ket .
\ee
\end{prop}

\proof 
Following (\ref{wkn2}), to get $\Phi_n(y)$ from $\{{\bf \Phi}^T_n(Y)\}_{T\in I_\pm^2}$ we need to restrict the sum in (\ref{gphy}) to $K=K_k$, $k\in\Z^d$, and to sum on all last steps $T\in H_\pm$. 

Let $\alpha$ be the distribution on $C^\infty(\T^{d}\times \T^{d})$ defined by
\be
\alpha(f(\cdot, \cdot))=\int_{\T^{d}} f(v, -v)d{\tilde v}.
\ee
Its Fourier coefficients satisfy $\hat \alpha (K)=\delta_{k,k'}$, for all $K=(k,k')\in \Z^{2d}$ so that
\be
\sum_{K=(k,k')}e^{iYK}W_K^T(n)\delta_{k,k'}=\alpha \star {\bf \Phi}^T_n(Y)
=\alpha({\bf \Phi}^T_n(Y-\cdot)).
\ee
On concludes using periodicity and by observing that the form $\bra {\bf \Psi_1}|$ yields the summation on $T\in H_\pm$. \ep

\begin{rem}
Noting that ${\bf \Psi}_1=\sum_{\tau\in I_\pm}|\tau\otimes\tau \ket$, we get
\bea\label{fitrace}
\Phi^{\ffi_0}_n(y)&=&\sum_{\tau\in I_\pm} \int_{\T^d} \bra\tau | \cJ_n(y-v)\ffi_0\ket\overline{\bra\tau |\cJ_n(-v)\ffi_0\ket} d{\tilde v}\nonumber\\
&=& \int_{\T^d} {\em \tr}(\cJ_n(y-v)|\ffi_0\ket\bra \ffi_0|\cJ^*_n(-v))\, d{\tilde v}\nonumber\\
&=& \int_{\T^d} {\em \tr}(\cJ^*_n(-v)\cJ_n(y-v)|\ffi_0\ket\bra \ffi_0|)\, d{\tilde v}.
\eea
\end{rem}

\bigskip

\subsection{Density matrices}

The analysis above can easily be adapted in order to accommodate more general initial vectors or density matrices.

A density matrix $\rho$ is a trace class non-negative operator on $\cH=\C^{2d}\otimes l^{2}(\Z^d)$ which can be represented by its kernel 
\be
\rho=(\rho(x,y))_{(x,y)\in\Z^d\times\Z^d}, \ \ \mbox{where }\ \rho(x,y)\in M_{2d}(\C) 
\ee
such that
\be
\rho=\sum_{(x, y)\in \Z^{2d}}\rho(x,y)\otimes |x\ket\bra y|.
\ee
The matrix $\rho(x,y)$ satisfies 
\be\label{symrho}
\rho(x,y)=\rho^*(y,x)\ \ \Rightarrow \ \rho(x,x)=\rho^*(x,x)\geq 0
\ee
 and  its elements are given by 
\be
\rho_{\sigma, \tau}(x,y), \ \ (\sigma,\tau)\in I_\pm^2, \ \ \ \mbox{so that} \ \bra \sigma\otimes x | \rho \, \tau\otimes y\ket=\rho_{\sigma, \tau}(x,y).
\ee
Since $\rho\geq 0$ is trace class, and $\C^d$ is finite dimensional,  we have
\bea\label{sumreg}
&&\sum_{x\in \Z^{d}}\tr \rho(x,x)=\|\rho\|_1<\infty, \\ 
&&\|\rho(x,x)\|\leq \tr \rho(x,x)\leq  2d \|\rho(x,x)\|. 
\eea

The expectation value of a lattice observable $F=\I\otimes f$ in the state corresponding to $\rho$ reads
\be
\bra F \ket_\rho=\tr (\rho (\I\otimes f))=\sum_{x\in\Z^d}f(x)\tr (\rho(x,x)),
\ee
where the first trace is on $\cH$ and the second on $\C^{2d}$, assuming that the sum converges.

If $\rho_0$ denotes the initial density matrix, its evolution at time $n$ under 
$U(n,0)$ defined by (\ref{evoln}) is given by
\be
\rho_n=U(n,0)\rho_0U^*(n,0)
\ee
and the expectation of the lattice observable $F$ is denoted by
\be
\bra F\ket_{\rho_0}(n)=\tr (\rho_n (\I\otimes f)),
\ee
if it exists.

Let us specify regularity properties on the lattice observable $F=\I\otimes f$ and the initial density matrix $\rho_0$ which imply that all manipulations below are legitimate.

\medskip
{\bf Assumption R:} \\
a) The lattice observable is such that, for any $\mu<\infty$, $\exists C_\mu<\infty$ such that
\be
|f(x+y)|\leq C_\mu |f(x)|, \ \ \forall \, (x,y)\in \Z^d\times\Z^d \ \ \mbox{with}\, \|y\|\leq\mu.
\ee
b) The kernel $\rho_0(x,y)$ is such that 
\bea
&&\sum_{(x,y)\in \Z^{d}\times\Z^d}\| \rho_0(x,y)\|<\infty \\
&&\sum_{x\in\Z^d}|f(x)|\|\rho_0(x,x)\|<\infty.
\eea

\medskip

In a similar fashion to Lemma \ref{expn}, we can express $\bra F\ket_{\rho_0}(n)$ in the following way
\begin{lem}\label{expnden} 
The kernel of $\rho_n$ reads
\be
\rho_n(x,y)=\sum_{(k,k')\in\Z^{d}\times\Z^d} J_k(n)\rho_0(x-k,y-k')J_{k'}^*(n).
\ee
Let $F=\I\otimes f$ and $\rho_0$ satisfy Assumption R. Then $\rho_n$ satisfies Assumption R b) and 
\bea
\bra F\ket_{\rho_0}(n)&=&\sum_{z\in\Z^d}f(z)\sum_{(k,k')\in\Z^{d}\times\Z^d}{\em \tr} (J_k(n)\rho_0(z-k,z-k')J_{k'}^*(n))\nonumber\\
&=&\sum_{z\in\Z^d}f(z)\sum_{(k,k')\in\Z^{d}\times\Z^d}{\em \tr} (J_{k'}^*(n)J_k(n)\rho_0(z-k,z-k')).
\eea
\end{lem}

\proof
Since for all $n \in \N$, the summations on $k$ and $k'$ are restricted to $\|k\|\leq \mu(n)$ and $\|J_k(n)\|\leq c(n)$,  we need to control
\bea\label{estimz}
\Sigma&:=&\sum_{z\in\Z^d}|f(z)|\sum_{(k, k') \in \Z^d\times \Z^d}
\|\rho_0(z-k,z-k')\|
\eea
under Assumption R.
Now,  $\rho_0\geq 0$ implies $P_{x y}\rho_0P_{x y}\geq 0$, where $P_{x y}$ is the orthogonal projector on $\cH$
\be
P_{x y}=\I \otimes |x\ket\bra x|+\I \otimes |x\ket\bra y|+\I \otimes |y\ket\bra x|+\I \otimes |y\ket\bra y|.
\ee
In other words, the following  $4d\times 4d$ block matrix is non-negative
\be
\begin{pmatrix}
\rho(x,x) & \rho(x,y) \\ \rho^*(x,y) & \rho(y,y)
\end{pmatrix}.
\ee
According to Lemma 1.21 in \cite{z}, this is equivalent to $\rho(x,x)\geq 0$, $\rho(y,y)\geq 0 $ and $\exists W$, $\|W\|\leq 1$ such that $\rho(x,y)=\rho(x,x)^{1/2}W\rho(y,y)^{1/2}$. Hence, 
\be
\|\rho(x,y)\|\leq \|\rho(x,x)\|^{1/2}\|\rho(y,y)\|^{1/2}.
\ee 

Applied to (\ref{estimz}), this yields together with Assumption R and Cauchy Schwarz,
\bea\nonumber
\Sigma&\leq& \sum_{z\in\Z^d}|f(z)|^{1/2}|f(z)|^{1/2}\sum_{(k, k') \in \Z^d\times \Z^d}
\|\rho_0(z-k,z-k)\|^{1/2}\|\rho_0(z-k',z-k')\|^{1/2}\nonumber\\
&\leq& C_{\mu(n)} \sum_{{(k, k') \in \Z^d\times \Z^d}\atop{\|k\|\leq \mu(n), \|k'\|\leq \mu(n)}}\sum_{z\in\Z^d}( |f(z-k)|\|\rho_0(z-k,z-k)\|)^{1/2}\times \nonumber\\
&&\quad \quad  \quad \quad  \quad \quad  \quad \quad  \quad \quad  \times (|f(z-k')|\|\rho_0(z-k',z-k')\|)^{1/2} \nonumber\\
&\leq& C_{\mu(n)} \left( 
\sum_{
\|k\|\leq \mu(n), \|k'\|\leq \mu(n)}1\right)\quad \sum_{x\in\Z^d}|f(x)|\|\rho_0(x,x)\|<\infty.
\eea
\ep

This generalization of Lemma \ref{expn} allows us to give an interpretation in terms of classical random walk on $\Z^d$
\bea
P(X_n=z)
&=&\tr (\rho_n(z,z))\equiv W^{\rho_0}_z(n),
\eea
with corresponding characteristic function $\Phi^{\rho_0}_n(y)=\sum_{z\in\Z^d}e^{iyz}W^{\rho_0}_z(n).$

Due to the expression of $W_z(n)$ as a convolution, the characteristic function will be expressed as a product of Fourier transforms.

\medskip

For $Y=(y,y')\in \T^d\times\T^d$, we define the matrix valued Fourier transform of a density matrix $\rho_0$ by
\be\label{defr}
R_0(Y)=\sum_{(k,k')\in\Z^{d}\times\Z^d} e^{i(yk+y'k')}\rho_0(k,k')\in M_{2d}(\C).
\ee
Because of (\ref{sumreg}) and (\ref{symrho}), $R_0$ is uniformly continuous in $Y$ and  satisfies
\be\label{symr}
R_0^*(y,y')=R_0(-y',-y).
\ee
Then, the Fourier  transform of $J_n$ being $\cJ$ Lemma \ref{Jfourier},
 the Fourier transform $R_n$ of $\rho_n$ reads
\bea
R_n(y,y')&=&\sum_{(x,x')\in\Z^{d}\times\Z^d} e^{i(yx+y'x')}\sum_{(k,k')\in\Z^{d}\times\Z^d} J_k(n)\rho_0(x-k,x'-k')J_{k'}^*(n)\nonumber\\
&=&\cJ_n(y)R_0(y,y')\cJ^*_n(-y')
\eea
Proceeding as above, we arrive at the generalization of (\ref{fitrace})
\begin{lem} For any $y\in \T^d$,
\bea\label{phider}
\Phi^{\rho_0}_n(y)&=&\int_{\T^d}  {\em \tr} \, R_n(y-v,v)\, d{\tilde v}\\ \nonumber
&=&\int_{\T^d}  {\em \tr} \left(  \cJ_n(y-v)R_0(y-v,v)\cJ^*_n(-v)\right)d{\tilde v}\\ \nonumber
&=&\int_{\T^d}  {\em \tr} \left(  \cJ^*_n(-v)\cJ_n(y-v)R_0(y-v,v)\right)d{\tilde v}.
\eea
\end{lem}

Let 
\be
{\bf R}_0(y,y')=\sum_{(\tau,\tau')\in I_\pm^2}\bra\tau |R_0(y,y')\tau'\ket\, |\tau\otimes\tau'\ket \in \C^{2d}\otimes \C^{2d}.
\ee
Then, making use of the identity
\be
M_n(y,y')M_{n-1}(y,y')\cdots M_1(y,y')=\cJ_n(y)\otimes \overline{\cJ_n(-y')},
\ee
it is straightforward to get from (\ref{phider}) the following generalization of Proposition \ref{phiphi}
\begin{prop}\label{phiden}
\be
\Phi^{\rho_0}_n(y)=\int_{\T^d}  \bra {\bf \Psi}_1| M_n(y-v,v)M_{n-1}(y-v,v)\cdots M_1(y-v,v){\bf R}_{0}(y-v, v)\ket d{\tilde v}.
\ee
\end{prop}
\begin{rem}
The map $y\mapsto {\bf R}_{0}(y-v, v)$ is continuous only under (\ref{sumreg}). Under Assumption R for an observable increasing at infinity, this map becomes more regular.
\end{rem}
\begin{rem}
The procedure consisting in extending the space $\C^{2d}$ where the $C_j$'s act to the tensor product $\C^{2d}\otimes\C^{2d}$ where we consider $C_j\otimes\overline C_j$ is parallel to the GNS construction. Once the Fourier transform in position space is taken, it allows us to write the action of the one-step dynamics in the coin variables as the action of a unitary matrix in $\C^{2d}\otimes\C^{2d}$ and to replace the density matrix by a vector.
\end{rem}

\section{Random framework}\label{rands}

For a deterministic non periodic set of coin operators, not much can be said about $\bra F \ket_{\psi_0}(n)$ in general. Therefore we consider the following random quantum dynamical system which defines a quantum walk with random update of the internal degrees of freedom at each time step. Let $C(\omega)$ be a random unitary matrix on $\C^{2d}$ with probability space $(\Omega,{\cal \sigma},d\mu)$, where $d\mu$ is a probability measure. We consider the random evolution operator obtained from sequences of i.i.d. coin matrices on $(\Omega^{\N^*},{\cal F},d\P)$, where ${\cal F}$ is the $\sigma$-algebra generated by cylinders and $d\P = \otimes_{k\in\N^*} d\mu$, by
\be\label{rdqs}
U_{\overline \omega}(n,0)=U_n(\overline \omega) U_{n-1}(\overline \omega)\cdots U_1(\overline \omega), \ \ \mbox{where} \ \ U_k(\overline \omega)=S\ (C(\omega_k)\otimes \I),
\ee
and $\overline \omega=(\omega_{1}, \omega_2, \omega_3, \dots)\in \Omega^{\N^*}$. The evolution operator at time $n$ is now given by a product of i.i.d. unitary operators on $\cH$.
We shall denote statistical expectation values with respect to $\P$ by $\E$. 

\medskip

All results of the previous section apply, with each occurrence of $C_s$ replaced by $C(\omega_s)$. A superscript $\overline\omega$ will mention the resulting randomness of the different quantities encountered. In particular, the random dynamical system at hands yields random matrices $J_k^{\overline \omega}(n)\in M_{2d}(\C)$, which, in turn, define random probability distributions $\{W_k^{\overline \omega}(n)\}_{n\in\N^*}$ on $\Z^d$ which satisfy (\ref{norm}) for all $n\in \N^*$ and $\overline\omega\in \Omega^{\N^*}$.  The corresponding characteristic functions $\Phi_n^{\overline \omega}$ become random Fourier series whereas ${\bf \Phi}_n^{\overline \omega}(Y)$ is obtained by the following product of i.i.d. random matrices
\be
{\bf \Phi}_n^{\overline \omega}(Y)=M_{\omega_n}(Y)M_{\omega_{n-1}}(Y)\cdots M_{\omega_1}(Y){\bf \Phi}_0
\ee
where, with $Y=(y,y')$,
\be
M_{\omega_s}(Y)=D(Y)V(\omega_s)=d(y)C(\omega_s)\otimes\overline{d(-y')C(\omega_s)}
\ee
are distributed according to the image of $d\mu$ by the inverse mapping $C\mapsto D(Y) C\otimes\overline{C}$.

\subsection{Diffusive averaged dynamics}

We consider in that section the statistical average of the motion performed by random quantum walk, and, more specifically, its diffusive characteristics. For the lattice observable $X^2$, we will derive results regarding the long time behavior of 
\be
\E(\bra X^2\ket_{\psi_0}^{\overline \omega})(n)=\E\bra U_{\overline \omega}^* (n,0)\psi_0| X^2 U_{\overline \omega}(n,0)\psi_0\ket=\E(\E_{W_k^{\overline \omega}(n)}(X_n^2)). 
\ee
It means, see (\ref{rvlat}), that we consider the motion corresponding to the averaged probability distributions defined by 
\be
w_k(n):=\E(W_k^{\overline \omega}(n)), \ \ k \in \Z^d, \ \ n\in \N^*,
\ee
with corresponding characteristic function
\be
\Phi_n(y)=\E_{w(n)}(e^{i yX_n})= \sum_{k\in\Z^d}w_k(n)e^{i ky}.
\ee
\begin{rem}
To stress the dependence on $\overline \omega$ in the distribution $W_k^{\overline \omega}(n)$, we shall denote the corresponding random vector on the lattice by $X_n^{\overline \omega}$.
When we consider the averaged distribution $w(n)$ instead, we shall write $X_n$ for the corresponding random vector. 
\end{rem}

Our results generalize those of \cite{KBH} in the sense that the distribution of the random coin matrices considered here is arbitrary. As a consequence, the analysis cannot be mapped to that of a persistent or correlated classical random walk on the lattice, as was observed in \cite{KBH}.
\medskip

Let $\cE$ and $\cM(Y)$ be the matrices  defined by
\be\label{defeM}
\cE=\E (V(\omega))=\E(C(\omega)\otimes \overline C(\omega)) \ \ \ \mbox{and} \ \ \cM(Y)=D(Y)\cE.
\ee
Note that while $V(\omega)$ is a unitary tensor product, its expectation $\cE$ is neither unitary, nor a tensor product in general. But $\|\cE\|\leq 1$ and therefore $\|\cM(Y)\|\leq 1$.
\medskip

Since the $\{C(\omega_s)\}_{s\in \N^*}$ are i.i.d., we immediately get from Corollary \ref{prodM} and Proposition \ref{phiphi} that 
\be
\E({\bf \Phi}^{\overline \omega}_n)(Y)=(\cM(Y))^n{\bf \Phi}_0,
\ee 
so that
\be\label{exchar}
\E (\Phi^{\overline \omega}_n)(y)=\int_{\T^d}\bra{\bf \Psi}_1 |(\cM(y-v,v))^n{\bf \Phi}_0\ket \, d{\tilde v}.
\ee
The analysis of the diffusive scaling limit (\ref{diffscal}) now relies on the spectral properties of the matrices $\cE$ and $\cM(Y)$.

\subsection{Spectral properties}\label{sp}

The structure of these matrices implies the following deterministic and averaged statements:
\begin{lem}\label{spel}
Let $V(\omega)=C(\omega)\otimes \overline C(\omega)$, $\cE=\E (V(\omega))$, $\cM(Y)=D(Y)\cE$  and let $\cS$ denotes the unitary involution defined by
$\cS \ffi\otimes\psi =\psi\otimes\ffi$, for all $\ffi, \psi\in \C^{2d}$. 

Then, for all $\omega$ and all $y$, 
{${\bf \Psi}_1=\sum_{\tau\in I_\pm}|\tau\otimes\tau\ket$ is invariant under $V(\omega), \cE, \cM(y,-y), \cS$ and their adjoints.} 
Consequently
\be
\|\cE\|={\em \spr}(\cE)=\|\cM(y,-y)\|={\em \spr}(\cM(y,-y))=1,
\ee
where ${\em \spr}$ denotes the spectral radius.
Moreover,
\be\label{sym}
\cS V(\omega)\cS= \overline V(\omega), \ \ \ \cS \cM(y,-y)\cS=\overline \cM(y,-y), \ \ \ \cS \cE \cS=\overline \cE,
\ee so that 
\be\label{realspec}
\sigma(V(\omega))=\overline{\sigma(V(\omega))}, \ \ \ \sigma(\cE)=\overline{\sigma(\cE)}, \ \ \ \sigma(\cM(y,-y))=\overline{\sigma(\cM(y,-y))}.
\ee
\end{lem}
\begin{rem}
For $\cM(Y)$, we only have
\be\label{boundspec}
{\em \spr}(\cM(Y))\leq \|\cM(Y)\| \leq 1.
\ee 
\end{rem}
\begin{rem} Before taking expectation values, $1$ is a eigenvalue at least $2d$ times degenerate for the unitary matrices $V(\omega)$ and $M_\omega(y,-y)$, for all $\omega$ and $y$, because of their tensor product structure (\ref{tpv}), (\ref{tpm}).
\end{rem}

\medskip

We shall work under an assumption which implies that in the long run, the averaged quantum walk loses track of interferences and acquires a universal diffusive behavior. At the spectral level, this is expressed by the fact that after taking expectation values, $1$ is the only eigenvalue of $\cM(y, -y)=\E(M_\omega(y,-y))$ on the unit circle and it is simple. 

Let $D(z,r)\subset\C$ denote the open disc of radius $r$ centered at $z\in\C$. 

\medskip

{\bf Assumption S: } 
For all $v\in [0,2\pi)^d=\T^d$,
\be
\sigma(\cM(-v,v))\cap \partial D(0,1)=\{ 1\} \ \ \mbox{and the eigenvalue $1$ is simple.}  
\ee
\begin{rem} Actually, because of the form of (\ref{exchar}), it is enough to consider the spectrum of the restriction of $\cM(-v, v)$ to the $\cM^*(Y)$-cyclic subspace for  generated by ${\bf \Psi}_1$. Set
 \be
 \cI = 
\mbox{\em Span } 
 \{{\cM ^*(Y)}^k{\bf \Psi}_1, \ k\in \N, \,  Y\in \T^d\times \T^d\}, 
 \ee
and let $P_{\cI}=P^*_{\cI}$ be the orthogonal projector onto $\cI$. If $P_\cI\neq \I$, 
we can work under the weaker 

\medskip

{\bf Assumption S': } 
For all $v\in [0,2\pi)^d=\T^d$,
\be
\sigma(\cM(-v,v)|_{\cI})\cap \partial D(0,1)=\{ 1\} \ \ \mbox{and the eigenvalue $1$ is simple.}  
\ee
 \end{rem}
 \medskip

Indeed, note that ${\cM^*(Y)}P_\cI=P_\cI {\cM^*(Y)}P_\cI $ so that, at the level of linear forms
\be
\bra {\bf \Psi}_1| {\cM (Y)}^n = \bra  {\cM^*(Y)}^nP_\cI {\bf \Psi}_1|= 
\bra {\bf \Psi}_1| (P_\cI {\cM (Y)}P_\cI)^n=\bra {\bf \Psi}_1| {\cM (Y)|_\cI}^n \,.
\ee
While it is often necessary in applications to use S', see the examples, we keep working under S below in order not to burden the notation.\\

Let ${\bf \Psi_0}={\bf \Psi_1}/\|{\bf \Psi_1}\|$. Under assumption S, ${\bf \Psi_0}$ spans the one dimensional spectral subspace of $\cM(-v,v)$ associated with the eigenvalue $1$. Moreover, by Lemma \ref{spel}, the corresponding rank one spectral projector reads $P=|{\bf \Psi_0}\ket\bra{\bf \Psi_0} |$ and is $v$-independent. With $Q=\I-P$, we have the spectral decomposition 
\be
\cM(-v,v)=P+Q\cM(-v,v)Q
\ee 
 where, under assumption S, $\exists \ \eps <1 $, independent of $v\in\T^d$ such that
$\spr Q\cM(-v,v)Q\leq \eps$. 

\medskip

In keeping with (\ref{exchar}) and the diffusive scaling (\ref{diffscal}) to be used below, we  perform a perturbative analysis of the spectrum of $\cM(y-v,v)$ for small values of $\|y\|$, uniformly in $v\in\T^d$. Let us introduce the following notation for $(y,v)\in \T^d\times\T^d$ 
\be
\cM_v(y)=\cM(y-v,v), \ \mbox{so that}\  \cM_v(0)=\cM(-v,v).
\ee
Now, with $D(y,y')=d(y)\otimes d(y')$, see (\ref{defd})
\bea
\cM_v(y)&=&D(y,0)\cM_v(0)=\cM_v(0)+\sum_{\tau\in I_\pm}(e^{iyr(\tau)}-1)|\tau\ket\bra\tau|\otimes\I\ \cM_v(0)\nonumber \\
&\equiv&  \cM_v(0) +F(y) \cM_v(0),
\eea
where $\|\cM_v(0)\|= 1$ and $\|F(y)\cM_v(0)\|\leq c \|y\|$, with $c$ independent of $v$.

Since the map $(y,v)\mapsto \cM_v(y)$ is actually analytic in $\C^d\times\C^d$, we can say more. For $\nu>0$, let  $\cT_\nu^d=\{z\in \C^d\ | \ \Re z \in \T^d, \ \Im z_j < \nu, j=1,\dots,d. \}\subset \C^d$ be a complex neighborhood of $\T^d$. For $y_0>0$, let $\cB(0,y_0)=\{y\in \C^d \ | \ \|y\|\leq y_0\}$.

Analytic perturbation theory, see \cite{k}, then yields the following 
\begin{lem}\label{specprop}
Under assumption S,  there exists $0<\delta<1$, $\nu=\nu(\delta)>0$ and  $y_0=y_0(\delta)>0$ such $(y,v)\in (\cT^d_\nu\cap \cB(0,y_0))\times \cT^d_\nu$  implies 
\bea
\sigma(\cM_v(y))\cap D(1,\delta)&=&\{\lambda_1(y,v)\}\, \\
\sigma(\cM_v(y))\setminus \{\lambda_1(y,v)\} &\subset& D(0,1-\delta).
\eea
Moreover, $\lambda_1(y,v)$ is simple, analytic in $(\cT^d_\nu\cap \cB(0,y_0))\times \cT^d_\nu$ and $\lambda_1(0,v)=1$ for all $v\in\cT_\nu^d$. The corresponding spectral decomposition reads
\be\label{specdec}
\cM_v(y)=\lambda_1(y,v)P(y,v)+{\cM_Q}(y,v),
\ee
where $P(y,v)$ is analytic in $(\cT^d_\nu\cap \cB(0,y_0))\times \cT^d_\nu$ and $P(0,v)=P$ for all $v\in\cT_\nu^d$. \\ With $Q(y,v)=\I -P(y,v)$, the restriction ${\cM_Q}(y,v)=Q(y,v)\cM_v(y)Q(y,v)$ satisfies ${\em \spr} (\cM_Q(y-v,v))<1-\delta$.
\end{lem}

We need to compute $\lambda_1(y,v)=\tr (P(y,v)\cM_v(y))$ to second order in $y$. We expand $F(y)$ as
\bea\label{fy3}
F(y)&=&F_1(y)+F_2(y)+O(\|y\|^3)\\ \nonumber
&=&\sum_{\tau\in I_\pm}iyr(\tau)\, |\tau\ket\bra\tau |\otimes \I - \sum_{\tau\in I_\pm}\frac{(yr(\tau))^2}{2}\, |\tau\ket\bra\tau | \otimes\I + O(\|y\|^3)
\eea
and introduce the (unperturbed) reduced resolvent $S_v(z)$
for $v\in\cT^d_\nu$ and $z$ in  a neighborhood of $1$ such that
\be\label{redres}
(\cM_v(0)-z)^{-1}=\frac{P}{1-z}+S_v(z)\ \ \mbox{with} \ P=|{\bf \Psi}_0\ket\bra {\bf \Psi}_0|.
\ee
We have for a simple eigenvalue, (see \cite{k} p.69)
\bea
\lambda_1(y,v)&=&1+\tr (F_1(y)\cM_v(0)P)\\ \nonumber 
&+&\tr ( F_2(y)\cM_v(0) P-F_1(y)\cM_v(0)S_v(1)F_1(y)\cM_v(0)P)+O_v(\|y\|^3).
\eea
Explicit computations with symmetry considerations yield
\begin{lem}\label{secord}
For all $v\in \cT^d_\nu$ and $y\in \cB(0,y_0)$, there exists a symmetric matrix $\D(v)\in M_{d}(\C)$  such that
\bea\label{difmat}
\lambda_1(y,v)&=&1+\frac{i}{2d}\sum_{\tau\in I_\pm}yr(\tau)+O_v(\|y\|^3)\nonumber \\
&&+\frac{1}{2d}\left(\sum_{\tau\in I_\pm}\frac{(yr(\tau))^2}{2}+\sum_{\tau,\tau'\in I_\pm}(yr(\tau))(yr(\tau'))\left\{\bra\tau\otimes\tau | S_v(1)\tau'\otimes\tau'\ket-\frac{1}{2d}\right\}\right)\nonumber \\
 &\equiv&1+\frac{i}{2d}\sum_{\tau\in I_\pm}yr(\tau)-\frac 12 \bra y | \D(v) y\ket+ O_v(\|y\|^3).
\eea
The map $v \mapsto \D(v)$ is analytic in $\cT_\nu^d$; when $v\in \T^d$, $\D(v)\in M_{d}(\R)$ is non-negative and $\D(v)_{j,k}=\frac{\partial^2}{\partial y_j \partial y_k}\lambda(0,v)$, $j,k \in \{1,2,\dots, d\}$. Moreover,  $O_v(\|y\|^3)$ is uniform in $v\in \cT_\nu^d$.
\end{lem}
\proof Existence and analyticity in $v$ of $\D(v)$ follow from analyticity of $\lambda_1$ in $y$ and analyticity of $S_v(1)$ in $v$, see (\ref{redres}). Since $\D(v)_{j,k}=\frac{\partial^2}{\partial y_j \partial y_k}\lambda(0,v)$,  the matrix is symmetric.
For $v\in\T^d$, the symmetry (\ref{sym}) implies $\cS S_v(1)\cS=\overline S_v(1)$ so that
\be
\bra\tau'\otimes\tau' | S_v(1)\tau\otimes\tau\ket = \bra\tau'\otimes\tau' | \cS S_v(1) \cS \tau\otimes\tau\ket = \overline{\bra\tau'\otimes\tau' | S_v(1)\tau\otimes\tau\ket}.
\ee
 Hence the matrix elements $\D(v)$ for $v\in\T^d$ are real as well. Finally, (\ref{boundspec}) implies that $\bra y | \D(v) y\ket\geq 0$ for all $y\in\T^d$. 
\ep

\begin{rem}
Using the notation  $\overline{f}=\frac{1}{2d}\sum_{\tau\in I_\pm}f(\tau)$ for any function on $I_\pm$, $\D(v)$ reads
\bea
\D(v)&=&2|\overline{r}\ket\bra \overline{r}|-\frac12\overline{\ |r\ket\bra r|\ }-\frac1d
\sum_{\tau,\tau'\in I_\pm}|r(\tau)\ket\bra\tau\otimes\tau | S_v(1)\tau'\otimes\tau'\ket\bra r(\tau')|,
\eea
where the "bra" and "ket" notation is understood in $\R^d$ for vectors $r(\tau)$ and in $\C^{2d}\otimes \C^{2d}$ for $\tau\otimes\tau $.
\end{rem}

\medskip

We are now set to prove the 
\begin{prop}\label{tecprop}
Under assumption S, uniformly in $v\in\cT^d_\nu$, in $y$ in compact sets of $\C$ and in $t$ in compact sets of $\R_+^*$,
\bea
&&\lim_{n\ra\infty}\cM_v^{[tn]}(y/{n})=e^{i{t} y \overline{r}}P,\\
&&\lim_{n\ra\infty}\cM_v^{[tn]}(y/\sqrt{n})e^{-i[tn]\overline{r}y/\sqrt{n}}=e^{-\frac{t}{2}\bra y | \D(v) y\ket}P.
\eea
\end{prop}
\proof Let $v\in\cT^d_\nu$, $t\in G\subset \R_+^*$, and $y\in K\subset \C$, $G$ and $K$  compact. We consider $n$ large enough so that  $y/\sqrt n$ and $y/n$ belong to $\cB(0,y_0)$, uniformly in $y\in K$. The decomposition (\ref{specdec}) implies for any $n$ large enough,
\be
\cM_v^{[tn]}(y)=\lambda_1^{[tn]}(y,v)P(y, v)+\cM^{[tn]}_Q(y, v)
\ee
where Lemma \ref{specprop} implies the existence of $c>0$ and $1>\delta'>\delta$, uniform in $(y,v,t)\in \cT^d_\nu\times K\times G$, such that 
\be
\|\cM^{[tn]}_Q(y, v)\|\leq c\delta'^{tn}.
\ee
Moreover, by Lemma \ref{secord},
\bea
\lambda_1^{[tn]}(y/n,v)&=&  \left( 1+i \frac{ \overline{r}y}{n}+O_v(\|y\|^2/n^{2}) \right)^{[tn]}\\ \nonumber
&{{ n\ra\infty} \atop \longrightarrow}& e^{it y  \overline{r}},\\
\lambda_1^{[tn]}(y/\sqrt n,v)e^{-i[tn]\frac{\overline{r}y}{\sqrt{n}}}&=&\left\{e^{-i\frac{\overline{r}y}{\sqrt{n}}}\left( 1+i\frac{\overline{r}y}{\sqrt{n}}- \frac{\bra y | \D(v) y\ket}{2n}+O_v\left(\frac{\|y\|^3}{n^{3/2}}\right) \right)\right\}^{[tn]}\\ \nonumber
&{n\ra\infty \atop \longrightarrow}& e^{-\frac{t}{2}\bra y | \D(v) y\ket}, \eea
and both $P(y/\sqrt n, v)$ and  $P(y/n, v)$ tend to $P$ as ${n\ra\infty }$, uniformly in $(v,y)\in  \cT^d_\nu\times K$. \ep
\medskip

With these technical results behind us, we come to the main results of this section which are the existence of a diffusion matrix  and central limit type behaviors. 

\medskip

Let  $\cN(0,\Sigma)$ denote the centered normal law in $\R^d$ with positive definite covariance matrix $\Sigma$ and let us write  $X^{\omega}\simeq \cN(0,\Sigma)$ a random vector $X^\omega\in\R^d$ with distribution $\cN(0,\Sigma)$. The superscript $\omega$ can be thought of as a vector in $\R^d$ such that for any Borel set $A\subset {\mathbb R}^d$
\be
\P(X^\omega\in A)=\frac{1}{(2\pi)^{d/2}\sqrt{\det (\Sigma)}}\int_{A}  e^{-\frac{1}{2}\bra \omega|\Sigma^{-1} \omega\ket} d\omega.
\ee
The corresponding characteristic function is $\Phi^{\cN}(y)=\E(e^{iyX^{\omega}})=e^{-\frac12\bra y | \Sigma y\ket}$.

\medskip 

The first result concerning the asymptotics of the random variable $X_n$ reads as follows.

\begin{thm} \label{cf}
Under Assumption S, uniformly in $y$ in compact sets of $\C$ and in $t$ in compact sets of $\R_+^*$,
\bea
&&\lim_{n\ra\infty}\Phi^{\ffi_0}_{[tn]}(y/ n)=e^{i{t} y  \overline{r}}\\
&&\lim_{n\ra\infty}e^{-i[tn]\frac{\overline{r}y}{\sqrt{n}}}\Phi^{\ffi_0}_{[tn]}(y/\sqrt n)=\int_{\T^d} e^{-\frac{t}{2}\bra y | \D(v) y\ket}\, d{\tilde v},
\eea
where the right hand side admits an analytic continuation in $(t,y)\in \C\times \C^2$.

In particular, for any $(i,j)\in \{1, 2, \dots, d\}^2$, 
\bea
&& \lim_{n\ra\infty}\frac{\bra X_i\ket_{\psi_0}(n)}{n}=\overline{r}_i \\
&&\lim_{n\ra\infty}\frac{\bra (X-n\overline{r})_i (X-n\overline{r})_j \ket_{\psi_0}(n)}{n}=\int_{\T^d} \D_{i\,j}(v)\, d{\tilde v}.
\eea
\end{thm} 
\begin{rem} 
We will call {\em diffusion matrices} both $\D(v)$ and 
$
\D=\int_{\T^d} \D(v)\, d{\tilde v}.
$

\label{more}
For any ${\bf s}=(s_1, s_2, \dots, s_d)\in \N^d$ with $|{\bf s}|=\sum_{j=1}^{d} s_j$ and $D^s_y=\left(\frac{\partial\phantom{x}}{\partial y_1}\right)^{s_1}\cdots \left(\frac{\partial\phantom{x}}{\partial y_d}\right)^{s_d}$,
\bea
&&\lim_{n\ra\infty }\frac{\bra X_1^{s_1}X_2^{s_2}\cdots X_d^{s_d}\ket_{\psi_0}(n)}{n^{|{\bf s}|}}=\overline{r}_1^{s_1}\overline{r}_2^{s_2}\cdots\overline{r}_d^{s_d},\\
&&\lim_{n\ra\infty }\frac{\bra (X-n\overline{r})_1^{s_1}(X-n\overline{r})_2^{s_2}\cdots (X-n\overline{r})_d^{s_d}\ket_{\psi_0}(n)}{n^{|{\bf s}|/2}}\nonumber\\
&&\quad\quad\quad\quad\quad\quad\quad\quad\quad\quad\quad\quad= (-i)^{|{\bf s}|}\int_{\T^d} (D^s_ye^{-\frac{1}{2}\bra y | \D(v) y\ket})|_{y=0}\, d{\tilde v},
\eea
which shows that all odd moments ($|{\bf s}|$ odd) of the centered variable are zero whereas all even moments can be computed explicitly. 
\end{rem}

\proof This is a direct consequence of Proposition \ref{tecprop} and definition (\ref{exchar}). The uniformity  of the convergence in the variables $(v,y,t)$  in compact sets  provides  analyticity after integration in $v\in \T^d$ and commutation of the limit and derivations. \ep

For initial conditions corresponding to a density matrix $\rho_0$, we have
\begin{cor}\label{cdiffeq}
Under Assumption S, for any $t\geq0$,
\bea
&&\lim_{n\ra\infty}\Phi^{\rho_0}_{[tn]}(y/n)=e^{i{t} y  \overline{r}}, \\
&&\lim_{n\ra\infty} e^{-i[tn]\frac{\overline{r}y}{\sqrt{n}}}\Phi^{\rho_0}_{[tn]}(y/\sqrt n)=\int_{\T^d} e^{-\frac{t}{2}\bra y | \D(v) y\ket}\, \bra {\bf \Psi}_1|{\bf R}_0(-v,v)\ket d{\tilde v} \nonumber\\
&&\hspace{4.4cm}=\int_{\T^d} e^{-\frac{t}{2}\bra y | \D(v) y\ket}\, {\em \tr} \,({ R}_0(-v,v)) d{\tilde v}, 
\eea
where 
\be
R_0(-v,v)=\sum_{(k,l)\in\Z^d\times\Z^d}e^{ivl}\rho_0(k,k+l).
\ee
\end{cor}
\begin{rem}
Under Assumption R for the observable $X^2$, we deduce that  for any $(i,j)\in \{1, 2, \dots, d\}^2$, 
\bea
&&\lim_{n\ra\infty}\frac{\bra X_i\ket_{\rho_0}(n)}{n}=\overline{r}_i,\\
&&\lim_{n\ra\infty}\frac{\bra (X-n\overline{r})_i(X-n\overline{r})_j\ket_{\rho_0}(n)}{n}=\int_{\T^d} \D_{i\,j}(v) {\em \tr} \,({ R}_0(-v,v)) \, d{\tilde v}.
\eea
\end{rem}
From Corollary \ref{cdiffeq}, and Theorem \ref{cf}, we gather that the characteristic function of the centered variable $X_n-n\overline r$ in the diffusive scaling $T=nt$, $Y=y/\sqrt{n}$, where $n\ra\infty$, converges to 
\be
\int_{\T^d} 
\cF\left(\frac{e^{-\frac{1}{2t}\bra x | \D^{-1}(v) x\ket} }{(t 2\pi)^{d/2}\sqrt{\det \D(v)}}\right)(y)
\, { \tr} \,({ R}_0(-v,v)) d{\tilde v},
\ee
where the function under the Fourier transform symbol $\cF$ is a solution to the diffusion equation
\be\label{diffusion}
\frac{\partial \ffi}{\partial t}=\frac12\sum_{i,j=1}^d\D_{ij}(v)\frac{\partial^2\ffi}{\partial_{x_i}\partial_{x_j}}.
\ee
As explained in \cite{ks}, \cite{hks}, it follows that the position space density $w_k([nt])\delta(\sqrt n x-k)$ converges in the sense of distributions to a superposition of solutions to the diffusion equations (\ref{diffusion}) as $n\ra\infty$.

\medskip

In case where the diffusion matrix $\D(v)=\D$ is independent of $v$,  Theorem \ref{cf} with $t=1$ says that the characteristic function of the rescaled variable $(X_n-n\overline r)/\sqrt n$ defined by  $\E_{w_k(n)}(e^{iy(X_n-n\overline r)/\sqrt n})$ converges to $e^{-\frac12\bra y|\D y\ket}$ which is the characteristic function of the normal law $\cN(0,\D)$. Hence, by L\'evy's continuity theorem, see {e.g. Theorem 7.6 in \cite{bil},
\begin{cor}
Assume S, suppose $\D(v)=\D>0$ is independent of $v\in \T^d$. Then, for any initial vector $\Psi_0=\ffi_0\otimes |0\ket$,  we have as $n\ra\infty$, in distribution,
 \bea
 \frac{X_n-n\overline r}{\sqrt n}&{\longrightarrow }&\ X^\omega\simeq \cN(0,\D).
 \eea
\end{cor}
\begin{rem}
All results of this section hold under Assumption S' only, {\it mutatis mutandis}. In particular, if the invariant subspace $\cI$ coincides with span $\{|\sigma\otimes\sigma\ket\}_{\sigma\in I_\pm}$, the matrix $\cM_v(y)$ is actually independent of $v$, because $D(-v,v)$ acts like the identity on the latter space. Consequently, the diffusion matrix is independent of $v$ as well. 
\end{rem}
\begin{rem}
We have chosen to randomize the coin state updates only, but it is possible to adapt the method to deal with random jump functions as well.  
\end{rem}

\subsection{Example}\label{ex1}

For $d=1$, consider the set of three unitary matrices in $\C^2$ given by $\left\{\un,\begin{pmatrix} 0& 1\\ 1&0 \end{pmatrix}, \frac{1}{{\sqrt2}}\begin{pmatrix} 1& 1\\ 1&-1 \end{pmatrix}\right\}$ and the distribution which assigns the probability $p/2>0$ to the first and second matrices and  $q=1-p$ to the third one. Let $r$ be the jump function defined by $r(\pm 1)=\pm 1$ so that $\overline r=0$. Then, in the ordered basis $\{|-1\otimes -1\ket, |1\otimes 1\ket, |-1\otimes 1\ket, |1\otimes -1\ket\}$, the corresponding matrix $\cE$ reads
\be
\cE=\frac12\begin{pmatrix}  1  & 1 & q & q  \\ 1  & 1 & -q & -q \\ q  & -q & p-q & 1 \\ q  & -q & 1 & p-q \end{pmatrix}
\ee 
and $D(-v,v)=\mbox{diag} (1,1, e^{i2v}, e^{-i2v})$. We introduce the following orthonormal basis whose first vector is ${\bf \Psi}_0$:
\bea
&& \{(|-1\otimes -1\ket +|1\otimes 1\ket)/\sqrt 2,  (|-1\otimes -1\ket -|1\otimes 1\ket)/\sqrt 2,|-1\otimes 1\ket, |1\otimes -1\ket\}\nonumber\\
&& \hspace{2cm } \equiv \{{\bf \Psi}_0, \ffi_1, \ffi_2, \ffi_3  \}.
\eea
In this basis, $\cM_v(0)=D(-v,v)\cE$ writes
\be \cM_v(0)=
\begin{pmatrix}  1  & 0 & 0 & 0 \\ 0  & 0 & \frac{q}{\sqrt2} & \frac{q}{\sqrt2}  \\ 0  & e^{i2v}\frac{q}{\sqrt2}  & e^{i2v}\frac{(p-q)}{2} & e^{i2v}\frac{1}{2} \\ 0  & e^{-i2v}\frac{q}{\sqrt2}& e^{-i2v}\frac{1}{2} & e^{-i2v}\frac{(p-q)}{2}  \end{pmatrix}\equiv1\oplus N_v,
\ee
where $N_v$ is the restriction of $\cM_v(0)$ to the subspace orthogonal to $\C {\bf \Psi}_0$. In order to make computations easier, we specialize to the case $p=1/\sqrt2$, so that 
\be
q/\sqrt2=(p-q)/2=(\sqrt2 -1)/2\equiv \gamma.
\ee
This way we can write in $(\C {\bf \Psi}_0)^\perp$
\be
N_v-\I=\begin{pmatrix}    -1 & \gamma & \gamma  \\  e^{i2v}\gamma  & e^{i2v}\gamma-1& e^{i2v}\frac{1}{2} \\  e^{-i2v}\gamma & e^{-i2v}\frac{1}{2} & e^{-i2v}\gamma-1  \end{pmatrix}.
\ee
We have $\det (N_v-\I)=2\cos(2v)(\gamma^2+\gamma)-(2\gamma^3+3/4)<0$ for all $v\in \T$ so that $\C{\bf \Psi}_0$ is the only invariant subspace under $\cM_v(0)$. Hence $S_v(1)=(N_v-\I)^{-1}$. To get the diffusion constant $\D(v)$ we need to compute $\bra \tau\otimes\tau |S(1) \tau'\otimes\tau'\ket$ for $\tau, \tau'=\pm1$, where $S(1)$ is defined on $(\C {\bf \Psi}_0)^\perp=Q\C^4$. We have $Q|\pm 1\otimes\pm 1\ket=\mp \ffi_1/\sqrt2$, so that
\be
\bra \tau\otimes\tau |S(1) \tau'\otimes\tau'\ket=\frac{\tau \tau'}{2} \bra \ffi_1 |(N_v-\I)^{-1}\ffi_1\ket=\frac{\tau \tau'}{2} \frac{\gamma^2-2\cos(2v)\gamma+3/4}{2\cos(2v)(\gamma^2+\gamma)-(2\gamma^3+3/4)}.
\ee
Taking into account the  $r(\tau)=\tau$ in formula (\ref{difmat}), we  get
\be
\D(v)y^2=-y^2(1+4 \bra \ffi_1 |(N_v-\I)^{-1}\ffi_1\ket),
\ee
which, plugging in the value of $\gamma$, eventually yields
\be\label{difex1}
\D(v)=\left(\frac{16-9\sqrt2+2\cos(2v)(5-4\sqrt2)}{5\sqrt2-4-2\cos(2v)}\right) >0.
\ee

\section{Einstein's Relation}\label{ein}

An interesting feature of the previous results is that the asymptotic averaged velocity $\overline{r}$ depends on the jump function $r$ only  and is independent of the coin distribution. This is reminiscent of the asymptotic velocity $v(F)$ reached by a particle subject to a deterministic force of amplitude $F$ in a random dissipative environment modeled by random forces. Given an asymptotic velocity, the mobility vector $\mu$ is defined as the ratio
\be\label{mu}
\mu=\lim_{F\ra 0}v_F/F.
\ee
This mobility $\mu$ is then related to the fluctuations of the system around the asymptotic trajectory by Einstein's relation which says that the diffusion matrix is proportional to $\|\mu\|$.

In the present framework, neither dissipation nor forces can be directly traced back to describe the asymptotic motion 
\be
\bra X\ket_{\Psi_0}(n)=n \overline{r}+o(n), \ \ \ n\ra\infty.
\ee
Moreover, the motion taking place on a lattice, the asymptotic velocity $\overline{r}$ has a minimal amplitude $1/(2d)$, if it is non zero, which prevents a behavior similar to (\ref{mu}).
Nevertheless, the jump function $r$ which characterizes the deterministic motion can be thought of as an external control parameter, similar to a driving force.

In order to get an asymptotic velocity which vanishes with the exterior control parameter, we rescale the lattice $\Z^d$ to $(\Z/l)^d$, with $l>0$. This means 
we consider the variable 
\be
Y_n=X_n/l \in  (\Z/l)^d.
\ee
Then we introduce a parameter $s\in \N$ as follows. Let $r_1$ and $r_0$ be two non-zero jump functions such that
\be
\overline{r_1}=0 \ \ \mbox{and} \ \ \ \overline{r_0}\neq 0.
\ee
We define a new $s$-dependent jump function by
\be\label{rho}
r_s(\tau)=s r_1(\tau)+ r_0(\tau)\in \Z^d\ \  \mbox{such that } \ \ \overline{r_s}=\overline{r_0}
\ee 
and we will consider the large $s$ limit. 
Hence the rescaled variable $Y_n$ satisfies
\bea
&&\lim_{n\ra \infty}\frac{\bra Y\ket_{\psi_0}(n)}{n}= \frac{\overline{r_0}}{l}:=v^{Y}\\
&&\lim_{n\ra\infty}\frac{\bra (Y-n\overline{r_s} )_i(Y-n\overline{r_s})_j \ket_{\psi_0}(n)}{n}=\frac{s^2}{l^2}\left(\int_{\T^d} {\D^{(1)}}_{i\,j}(v)\, d{\tilde v}+O(1/s)\right):=\D^Y_{i\,j}
\eea
where $\D^{(1)}(v)$ is the diffusion matrix computed by means of the jump function $r_1$ and the remainder term is uniform in $v\in \T^d$. Therefore, choosing the scale $l=s\in\N$, we get that the diffusion matrix $\D^Y$ is finite for large $s$ whereas the asymptotic velocity tends to zero. Hence, setting $F=1/s$,  we get for $s$ large
\bea
\mu&=&\lim_{F=1/s\ra 0}v^{Y}/F= \overline{r_0}   \\
\D^{Y}&=&\int_{\T^d} {\D^{(1)}}(v)\, d{\tilde v}+o(1/s)\ra K\|\mu\|, \ \ \mbox{as} \ \ \ s\ra \infty.
\eea
The last formula is admittedly a consequence of a rather ad hoc construction. 
On the other hand, assuming that $\overline{r_1}\neq 0$, we get with the same scaling $v^Y=\overline{r_1}$ which never vanishes.

\section{Moderate Deviations}\label{moder}

The spectral properties of the matrix $\cM(y-v,v)$ proven in Section \ref{sp} allow us to obtain further results on the behavior with $n$ of the distribution of the random variable $X_n$ defined by (\ref{defx}). This section is devoted to establishing some moderate deviations results on the random variable $X_n$.

We consider initial conditions of the form $\psi_0=\ffi_0\otimes |0\ket$ and we will be concerned with $X_n-n\overline r$. Moderate deviations results depend on asymptotic behaviors in different regimes of the logarithmic generating function of $X_n-n\overline r$ defined for $y\in\R^d$ by
\be
\Lambda_n(y)=\ln(\E_{w(n)}(e^{y(X_n-n\overline r)}))\in (-\infty, \infty].
\ee
This function $\Lambda_n$ is convex and $\Lambda_n(0)=0$.

Let $\{a_n\}_{n\in \N}$ be a positive valued sequence such that 
\be\label{an}
\lim_{n\ra\infty}a_n=\infty, \ \mbox{ and } \ \ \lim_{n\ra\infty}a_n/n=0.
\ee Define $Y_n=(X_n-n\overline r)/\sqrt{n a_n}$ and, for any $y\in \R^d$, let $\tilde \Lambda_n(y)=\ln(\E_{w(n)}(e^{yY_n}))$ be the logarithmic generating function of $Y_n$. 

\begin{prop} Assume S and further suppose $\D(v)>0$ for  all $v\in \T^d$. Let $y\in \R^d\setminus \{0\}$ and assume the real analytic map  $\T^d\ni v\mapsto \bra y | \D(v)y\ket\in\R_*^+$ is either constant or admits a finite set $\{v_j(y)\}_{j=1, \cdots, J}$ of non-degenerate maximum points in $\T^d$.
Then, for any $y\in\R^d$, 
\bea
\lim_{n\ra\infty}\frac{1}{a_n}\tilde \Lambda_n(a_n y)=\frac12 \bra y | \D(v_1(y))y\ket
\eea
which is a smooth convex function of $y$.
\end{prop}
\proof 
This proposition essentially follows from Lemmas \ref{specprop} and \ref{secord} and the asymptotic evaluation of an integral. Let $b_n=\sqrt{a_n/n}$ s.t. $\lim_{n\ra\infty}b_n=0$. By construction, $\tilde \Lambda_n(a_n y)=\Lambda_n(b_n y)$ where, according to Lemma \ref{specprop} and \ref{secord}, there exists $\gamma >0$ s.t. for $n$ large enough,
\bea\label{dd}
\exp(\Lambda_n(b_n y))&=& \int_{\T^d} \left(1+\frac{b_n^2}{2}\bra y | \D(v)y\ket+ O_v(b_n^3y^3)\right)^n \left(1+O_v(b_ny)\right)  \, d\tilde v + O(e^{-\gamma n})\nonumber\\
&=&\int_{\T^d} e^{\frac{a_n}{2}\bra y | \D(v)y\ket+O_v(a_n b_ny^3)} \left(1+O_v(b_ny)\right)  \, d\tilde v + O(e^{-\gamma n}).
\eea
All remainder terms $O_v(\cdots)$ are analytic in $v\in \cT_\nu$, as well as $\D(v)$. An application of Laplace's  method around each of the non-degenerate maximum points, yields for $1/3<\alpha<1/2$
\bea
\exp(\Lambda_n(b_n y))&=&\sum_{j=1}^Je^{\frac{a_n}{2}\left(\bra y | \D(v_j(y))y\ket+ O( b_ny^3)\right)}(G_j(y)/a_n^{d/2}+O(b_ny) +O(1/a_n^{3\alpha -1}) )\nonumber\\
& &+ O(e^{-\gamma n})+O(e^{-Ka_n^{1-2\alpha}}),
\eea
where $G_j(y)>0$, $K>0$, from which the result follows. The case where $\D(v)$ is independent of $v$ follows directly from (\ref{dd}). The convexity of the limit follows from the convexity of $\tilde \Lambda_n$.  The assumed non-degeneracy of the maximum point ensures that the functions $\R^d\setminus \{0\}\ni y\mapsto v_j(y)$ are all smooth by the implicit function theorem.\ep\\

Let us recall a few definitions notations. A {\em rate function} $I$ is a lower semicontinuous map from $\R^d$ to $[0,\infty]$ s.t. for all $\alpha\geq 0$, the level sets $\{x\ | \ I(x)\leq \alpha\}$ are closed. When the level sets are compact, the rate function $I$ is called {\em good}. For any $\Gamma\subset \R^d$, $\Gamma^0$ denotes the interior of $\Gamma$, while $\overline{\Gamma}$ denotes its closure.

As a direct consequence of G\"artner-Ellis Theorem, see \cite{dz} Section 2.3, we get
\begin{thm}\label{md} Define $\Lambda^*(x)=\sup_{y\in \R^d}\left(\bra y|x\ket -\frac12\bra y | \D(v_1(y))y\ket\right)$, for all $x\in \R^d$. Then, $\Lambda^*$ is a good rate function and, any positive valued sequence $\{a_n\}_{n\in \N}$ satisfying (\ref{an}) and  all $\Gamma\subset \R^d$
\bea
-\inf_{x\in \Gamma^0}\Lambda^*(x)&\leq &\liminf_{n\ra\infty}\frac{1}{a_n}\ln (\P((X_n-n\overline r)\in \sqrt{na_n}\, \Gamma))\nonumber\\
&\leq&\limsup \frac{1}{a_n}\ln (\P((X_n-n\overline r)\in \sqrt{na_n}\, \Gamma))\leq -\inf_{x\in \overline{\Gamma}}\Lambda^*(x).
\eea
\end{thm}
\begin{rem} As a particular case, when $\D(v)=\D>0$ is constant, we get
\be
\Lambda^*(x)=\frac12\bra x| \D^{-1}x\ket.
\ee
\end{rem}
\begin{rem}
Specializing the sequence $\{a_n\}_{n\in \N}$ to a power law, {\it i.e.} taking $a_n=n^\alpha$, we can express the content of Theorem \ref{md} in an informal way as follows. For $0<\alpha<1$, 
\be
\P((X_n-n\overline r)\in n^{(\alpha +1)/2}\, \Gamma)\simeq e^{-n^\alpha \inf_{x\in {\Gamma}}\Lambda^*(x)}.
\ee
For $\alpha$ close to zero, we get results compatible with the Central Limit Theorem and for $\alpha$ close to one, we get results compatible with those obtained from a large deviation principle.
\end{rem}

\bigskip

Let us come back to the example in section \ref{ex1}. The diffusion coefficient $\D(v)$ given in (\ref{difex1}) admits , as a function of $v\in\T$, a single non-degenerate maximum at $v=0$ where it takes the value $\D(0)=2\sqrt2-1$. Thus we get from the foregoing that a moderate deviation principle holds for this example, with the good rate function $\Lambda^*(x)=\frac{x^2}{2\D(0)}=\frac{x^2}{2(2\sqrt 2 -1)}$.

\section{Example of diffusive  random dynamics}\label{four}

The results obtained so far can be viewed, essentially, as an adaptation to the quantum walk dynamics setup of those proven in \cite{p}, \cite{ks}, \cite{hks} for the averaged dynamics and as an extension of \cite{KBH},  \cite{jm}.

\medskip

In this section we consider a specific example of measure $d\mu$ on $U(2d)$, the set of coin matrices, for which we can prove convergence results on the associated random quantum dynamical system (\ref{rdqs}) for large times, in distribution rather than in average. In particular, our example shows that almost sure convergence results cannot be expected in general. 

As noted in Section \ref{sp}, the spectra of $V(\omega)$ and $M_{\omega}(y,-y)$ lying on the unit circle and admitting $1$ as a $2d$-fold eigenvalue prevent us from using the same spectral methods as above. For the same reason, the results about products of random contractions in \cite{bjm} do not  apply. 
However, the structure of the example at hand allows for a direct approach which, eventually, reduces the analysis to that of a central limit theorem for a Markov chain.

\subsection{Permutation matrices}

Let ${\mathfrak S}_{2d}$ be the set of permutations of the $2d$ elements of $I_\pm=\{\pm 1, \pm 2, \dots, \pm d\}$. For $\pi\in {\mathfrak S}_{2d}$ and $\Theta=\{\theta_j \}_{j\in I_\pm}\in \T^{2d}$,  define 
\be\label{defmatper}
C(\pi,\Theta)=\sum_{\tau\in I_\pm}e^{i\theta_{\pi(\tau)}}|\pi(\tau)\ket\bra \tau | \in U(2d) \ \ \mbox{so that }\ C_{\sigma \tau}(\pi,\Theta)=e^{i\theta_{\sigma}}\delta_{\sigma, \pi(\tau)}.
\ee
With $\Delta(\Theta)=\mbox{diag\,}(e^{i\theta})$ and $C(\pi)\equiv C(\pi, 0)$, we can write
\be
C(\pi,\Theta)=\Delta(\Theta)C(\pi), 
\ee
where $C(\pi)$ is a permutation matrix associated with $\pi$. We recall the following elementary properties: 
For any $\pi, \sigma \in {\mathfrak S}_{2d}$, 
\bea
C(\I)=\I, \ \ C^*(\pi)=C^T(\pi)=C(\pi^{-1}), \ \ C(\pi)C(\sigma)=C(\pi\sigma).
\eea
Moreover, Birkhoff-Von Neumann Theorem asserts that the set of doubly stochastic matrices of order $n$ is the convex hull of the set of permutation matrices of order $n$ whose extreme points coincide with the permutation matrices.

\medskip

The matrices $C(\pi, \Theta)$ allow for explicit computations of the relevant quantities introduced in Section \ref{dets}. It is easy to derive the next
\begin{lem}\label{lj} Let $r:I_\pm\ra \Z^d$ be a jump function. 
Given a sequence of $n$ permutations $\pi_1, \pi_2, \dots, \pi_n$, let
$(\tau_1, \tau_2, \dots, \tau_n)\in I_\pm^n$ be the sequence parametrized by $\tau_1$ given by $(\tau_1, \pi_2(\tau_1), \pi_3(\tau_2), \dots, \pi_{n}(\tau_{n-1}))$, {\em i.e.} such that
\be\label{path}
\tau_j=(\pi_j\pi_{j-1}\cdots\pi_2)(\tau_1), \ j=2, \dots, n.
\ee
Let $\Theta_1, \Theta_2, \dots, \Theta_n$ be a set of phases, $\Theta_j=(\theta_1(j), \dots, \theta_n(j))$.   
Then, with the convention $C_j=C(\pi_j, \Theta_j)$, we get for all $k\in\Z^d$, 
\bea
J_{k}(n)&=&\sum_{\tau_1\in I_\pm \ s.t. \atop  \sum_{j=1}^nr(\tau_j)=k}e^{i(\theta_{\tau_1}+\cdots + \theta_{\tau_n})}|\tau_n\ket\bra \pi_1^{-1}(\tau_1)|,
\eea
and $J_k(n)=0$, if $\sum_{j=1}^nr(\tau_j)\neq k$.
\end{lem}
Consequently, the non-zero probabilities $W_k(n)$ on $\Z^d$ read for any normalized internal state vector $\ffi_0$ and any density matrix $\rho_0$
\bea
W_k^{\ffi_0}(n)&=&\|J_k(n)\ffi_0\|^2= \sum_{\tau_1\in I_\pm \ s.t. \atop  \sum_{j=1}^nr(\tau_j)=k}|\bra \pi_1^{-1}(\tau_1)|\ffi_0\ket|^2 ,\\ \nonumber 
W_k^{\rho_0}(n)&=&{ \tr} \rho_n(k,k)=\sum_{j\in\Z^d}\sum_{\tau_1\in I_\pm\atop j=\sum_{s=1}^nr(\tau_s) }\bra \pi_1^{-1}(\tau_1)|\rho_0(k-j,k-j)\pi_1^{-1}(\tau_1)\ket .
\eea
Note that the sets of phases $\Theta_j$, $j=1, \dots, n$ play no role in the computation of expectation values of lattice observables. We set $\tau_1=\pi_1(\tau_0)$ and note 
\be
\ffi_0=\sum_{\tau_0 \in I_\pm}a_{\tau_0} |\tau_0\ket \ \ \Rightarrow \ \ |\bra \pi_1^{-1}(\tau_1)|\ffi_0\ket|^2=\sum_{\tau_0\in\I_\pm}|a_{\tau_0}|^2\delta_{\tau_1, \pi_1(\tau_0)}.
\ee
Hence $W_k^{\ffi_0}(n)= \sum_{\tau_0\in I_\pm }
|a_{\tau_0}|^2\delta_{\sum_{j=1}^nr(\tau_j), k}$ \ so that for $F=\I\otimes f$ and $\psi_0=\ffi_0\otimes |0\ket\bra 0|$
\be
\bra F\ket_{\psi_0}(n)= \sum_{k\in \Z^d}W_k^{\ffi_0}(n)f(k)= \sum_{\tau_0\in I_\pm }|a_{\tau_0}|^2 f(\sum_{j=1}^nr(\tau_j)).
\ee
\begin{rem}
In other words, given a set of $n$ permutations, there is no more quantum randomness in the variable $X_n$, except in the initial state. 
\end{rem}
Therefore the characteristic functions take the form
\begin{cor} With $\tau_j=(\pi_j \pi_{j-1}\cdots \pi_1)(\tau_0)$, for $j=1,\dots, n$,
\bea
\Phi^{\ffi_0}_n(y)&=&\sum_{\tau_0\in I_\pm }e^{iy \sum_{j=1}^nr(\tau_j)}|a_{\tau_0}|^2,  \\
\Phi^{\rho_0}_n(y)&=&\sum_{\tau_0\in I_\pm }e^{iy \sum_{j=1}^nr(\tau_j)}\int_{\T^d} 
\bra \tau_0| R_0(y-v,v) \tau_0\ket
\, d \tilde v.
\eea
\end{cor}
The dynamical information is contained in the sum $S_n=\sum_{j=1}^nr(\tau_j)$ which appears in the phase. The next section is devoted to its study, in the random version of this model where the coin matrices are  i.i.d. random variables with values in   $\{C(\pi, \Theta), \, \pi\in {\mathfrak S}_{2d}, \, \Theta\in \Z^{2d} \}$.

\subsection{Random dynamics}

Assume a random variable $C(\omega)$ with values in $\{C(\pi, \Theta)\in U(2d), \ (\pi, \Theta)\in  {\mathfrak S}_{2d}\times\T^{2d}\}$ is defined on a probability space $(\Omega, \sigma, d\nu)$. The foregoing shows that only the marginal $\alpha$ defined on the discrete set $\{ C(\pi)\in U(2d), \pi\in { \mathfrak S}_{2d}\}$ or, equivalently on $\{ \pi\in { \mathfrak S}_{2d} \}$, matters
\be\label{mmm}
\mu(\pi)\equiv\mu(\alpha(\omega)=\pi)=\nu(\{C(\omega)=C(\pi, \Theta)\, | \, \Theta\in \T^{2d}\}), \ \ \pi\in { \mathfrak S}_{2d}.
\ee 
We shall use the notation  $\alpha(\omega)\equiv \omega\in  { \mathfrak S}_{2d}$ and $\Omega={ \mathfrak S}_{2d}$.  The corresponding process is denoted by
$\overline \omega=(\omega_1, \omega_2, \omega_3, \dots )\in \Omega^{\N^*}$ and $d\P=\otimes_{k\in\N^*}d\mu$.

\medskip

Given $\ffi_0\in \C^{2d}$ an initial internal state and a random sequence of permutation matrices $(\omega_1,  \dots, \omega_n )$, the random variable $S_n(\overline \omega)=\sum_{j=1}^n\tau_j(\overline\omega)\in \Z^d$ is the sum of random variables
$\tau_j(\overline \omega)$, $j=1,\dots,n$ whose properties are given in the next lemma:
\begin{lem} Let $\ffi_0=\sum_{\tau_0}a_{\tau_0}|\tau_0\ket$ be the initial condition.
The path $(\tau_0, \tau_1, \dots, \tau_n)$ is a Markov chain with finite state space $I_\pm$ characterized by the initial probability distribution
\bea\nonumber
p_0(\tau_0=\sigma_0)&=&|a_{\sigma_0}|^2
\eea
and by the stationary transition probabilities 
\bea
P(\sigma', \sigma)=\em{\mbox{Prob}}(\tau_k(\overline \omega)=\sigma | \tau_{k-1}(\overline \omega)=\sigma'), &&  k=2,3, \dots, n
\eea
given by
\bea
P(\sigma', \sigma)&=&\sum_{\pi \in { \mathfrak S}_{2d}}\mu(\pi )\delta_{\sigma, \pi(\sigma')}. 
\eea
The corresponding transition matrix $P=(P(\sigma', \sigma))\in M_{2d}(\R^+)$ is doubly stochastic  and
\be
P=\E(C^T(\omega)).
\ee
\end{lem}
\begin{rem}
The transition matrix  $P$ is unitary iff $\mu(\pi)=\delta_{\pi_0, \pi}$ for some $\pi_0$.
\end{rem}
\proof 
By Lemma \ref{lj}, $\tau_k(\overline \omega)$ only depends on $\{\omega_j\}_{j=1,\cdots, k}$ and $\tau_0$ and is given by $\tau_k(\overline \omega)=\omega_k(\tau_{k-1})$. Hence
\bea
\mbox{Prob}(\tau_k(\overline \omega)=\sigma | \tau_{k-1}(\overline \omega)=\sigma')&=&\mbox{Prob}(\omega_k(\tau_{k-1})=\sigma | \tau_{k-1}=\sigma')\\ \nonumber
 &=&\mu(\{\omega \, | \, \omega(\sigma')=\sigma\})=\sum_{\pi \in {\mathfrak S}_{2d}}\mu(\pi)\delta_{\sigma, \pi(\sigma')},
\eea
where we used the independence of the $\omega_k$. Finally, (\ref{defmatper}) shows that the right hand side is the expectation of $C^T(\omega)$ w.r.t $\mu$.
\ep\\

Considering 
the diffusive scaling (\ref{diffscal}), we are thus naturally lead to investigate the large $n$ behavior of the quantity
\bea
\frac{1}{\sqrt{n}} S_n({\overline \omega})&=&\frac{1}{\sqrt{n}}\sum_{j=1}^n r(\tau_j(\overline \omega)),
\eea 
{\em i.e.} to a functional central limit theorem for the Markov chain  $(\tau_0, \tau_1, \tau_2, \dots )$ with finite state space $I_\pm$, initial probability $p_0$ and transition matrix $P$. 
\medskip

There are simple conditions under which a functional central limit theorem holds for Markov chains with finite state space, see {\em e.g.} 
in \cite{bil}. Let us recall the few basic notions and results associated to Markov chains with finite state space, $F$, characterized by a transition matrix $P\in M_{|F|}(\R_+)$ s.t. $\sum_{\tau\in F}P(\sigma, \tau)=1$ that we will need below. \bigskip

 A transition matrix $P$ is  {\em irreducible} if,  $\forall \, \sigma, \tau \in F$, $\exists \, n \in \N^*$ such that $P^n(\sigma, \tau)>0$. 
A probability distribution $p_0$, considered as a vector in $\R^{|F|}$, is  {\em invariant} for the transition matrix $P$ if $P^Tp_0=p_0$.
If $p_0$ is invariant, then the Markov process is stationary,
\be
{ \mbox{Prob}}((\tau_0,\tau_1,\cdots)\in B)={ \mbox{Prob}}((\tau_k,\tau_{k+1},\cdots)\in B), \ \ \forall \, k\in \N, \, B\subset F^{\N}.
\ee
If $P$ is irreducible, the invariant distribution $p_0$ is unique and  $p_0(\tau)>0,  \forall  \tau\in F$.
If $P$ is furthermore doubly stochastic, the invariant distribution $u_0$ is uniform $u_0(\tau)=1/|F|, \forall \tau\in F$.\\

In terms of spectral properties, an irreducible stochastic matrix $P$ admits $1$ as a simple eigenvalue. If it is furthermore doubly stochastic, the uniform vector $u_0$ is invariant under both $P$ and $P^*$.

\bigskip

Hence, if  we  take as initial distribution the uniform measure $u_0(\tau)=1/(2d)$ $\forall \, \tau\in I_\pm$, which is invariant for the doubly stochastic transition matrix $P=\E_\mu(C^T(\omega))$, the Markov process is stationary. Moreover,
\be\label{e0}
\E_{u_0}(r(\tau_0))=\frac{1}{2d}\sum_{\tau\in I_\pm}r(\tau)=\overline{r}.
\ee

 From  Thm 20.1 in \cite{bil} and its applications page 177, or \cite{l}, we have
\begin{thm}\label{clt}
Let $\ffi_0=\sum_{\tau\in I_\pm}a_\tau|\tau\ket$ and $p_0$ s.t. $p_0(\tau)=|a_\tau|^2$. Assume the transition matrix $P=\E(C^T(\omega))$ is irreducible. Then, 
$
\lim_{n\ra\infty}\frac{1}{ n} \sum_{j=1}^{n} r(\tau_j(\overline \omega))=\overline r
$
almost surely 
and, with convergence in distribution,
 \bea
\frac{1}{\sqrt n} \sum_{j=1}^{n} \left(r(\tau_j(\overline \omega))-\overline{r}\right) &{ { n\ra\infty} \atop \longrightarrow}& X^\omega\simeq \cN(0,\Sigma),
 \eea
provided the covariance matrix 
\bea\label{covma}
\Sigma_{ij}&=&-\frac{1}{2d}\bra r_i |r_j\ket+\overline r_i \overline r_j-\frac{1}{2d}\left(
 \bra r_i | S(1)r_j \ket+  \bra r_j | S(1)r_i \ket\right )
\eea
is definite positive, where $S(1)$ denotes the reduced resolvent of $P$ at $1$.
\end{thm}

\begin{rem}
An alternative formulation for $\Sigma$ is
\bea
\Sigma_{ij}&=&\frac{1}{4d}\left(\bra r_i | (\un_Q-P_Q )^{-1}(\un_Q+P_Q)r_j\ket+
 \bra r_j | (\un_Q-P_Q )^{-1}(\un_Q+P_Q) r_i\ket \right ),
\eea
where the projector $Q$ and the operator $P_Q$ are defined in the spectral decomposition of $P$ 
\be\label{sdp}
P={2d}|u_0\ket\bra u_0|+P_Q, \ \ \mbox{with }\ P_Q=QPQ \ \  \mbox{and}\ Q=Q^2=Q^*=\un-{2d}|u_0\ket\bra u_0|.
\ee

\end{rem}

\proof Our assumptions imply that $P$ is irreducible, doubly stochastic and that $u_0$ is invariant for $P$ and $P^*$. This together with (\ref{e0}) allows us to apply Thm 20.1 of \cite{bil} and the remarks p.177 or the results of \cite{l}.
It remains to compute the covariance matrix. Let us define the centered random vector
\be
\tilde r(\tau(\overline\omega))=r(\tau(\overline \omega))-\overline r
\ee
such that $\E_{u_0}(\tilde r(\tau_0))=0$, where  $\E_{u_0}$ denotes the expectation with invariant initial measure $u_0$. 
The first mentioned reference yields the following expression for the covariance matrix
 \be
 \Sigma_{ij}=\E_{u_0}(\tilde r(\tau_0)_{i}\tilde r(\tau_0)_{j})+\sum_{k=1}^\infty\E_{u_0} (\tilde r(\tau_0)_{i}\tilde r(\tau_k(\overline \omega))_{j}+\tilde r(\tau_k(\overline \omega))_{i}\tilde r(\tau_0)_{j}),
 \ee
for $ i, j = 1, 2, \dots, d,$ where $\tilde r(\tau)_{j}$ denotes the $j^{\mbox{\tiny th}}$ component of $\tilde r(\tau)\in \Z^d$.
We  compute  for any $k\in\N$
\bea\label{expec}
\E_{u_0}(\tilde r(\tau_0)_{i}\tilde r(\tau_k)_{j})&=&\frac{1}{2d}
\bra r_i | P^kr_j\ket- \overline r_i \overline r_j .
\eea
Note that the right hand side of (\ref{expec}) is equal to
\be\label{nnn}
\frac{1}{2d}\left(
\bra r_i | P^k(r_j -2d\, \overline r_ju_0)\ket\right) \ \ \ \mbox{with} \ \ \ \bra u_0|
(r_j -2d\, \overline r_ju_0)\ket=0.
\ee
By (\ref{sdp}),
for any $v, w \in \C^{2d}$,
\be
v -2d\, \overline vu_0=Q v \ \ \mbox{and } \ \ \bra w|Qv\ket=\bra w|v\ket-2d \overline w \,\overline v.  
\ee
Thanks to (\ref{nnn}), we can write
\be
\sum_{k=1}^\infty P^k(r_j -2d\, \overline r_ju_0)\ket=\sum_{k=1}^\infty P^k_Q (r_j -2d\, \overline r_ju_0)\ket=(\un_Q - P_Q)^{-1}P_Q(r_j -2d\, \overline r_ju_0),
\ee
where $\un_Q$ is the identity reduced to the subspace $Q\C^{2d}$. Therefore,  
\be
(\un_Q - P_Q)^{-1}P_Q\equiv -(S(1)+\un_Q) , 
\ee
where $S(1)=QS(1)Q=(P_Q-\un_Q )|_{Q\C^{2d}}^{-1}$ denotes the reduced resolvent of $P$ at $1$. Hence 
\bea
 \Sigma_{ij}&=&\frac{1}{2d}\bra r_i | r_j\ket - \overline r_i \overline r_j-\frac{1}{2d}(\bra r_i | (S(1)+\un_Q)(r_j -2d\, \overline r_ju_0)\ket \nonumber\\
 & & \hspace{5.5cm}+\bra r_j | (S(1)+\un_Q)(r_i -2d\, \overline r_iu_0)\ket)
 \nonumber\\
 &=&-\frac{1}{2d}\bra r_i |Q r_j\ket-\frac{1}{2d}\left(
 \bra r_i | S(1)r_j \ket+  \bra r_j | S(1)r_i \ket\right )
 \nonumber\\
 &=&-\frac{1}{4d}\left(\bra r_i | (Q+2S(1)) r_j\ket+\bra r_j | (Q+2S(1)) r_i\ket
 \right ) \nonumber\\
 &=&\frac{1}{4d}\left(\bra r_i | (\un_Q-P_Q )^{-1}(\un_Q+P_Q)r_j\ket+
 \bra r_j | (\un_Q-P_Q )^{-1}(\un_Q+P_Q) r_i\ket \right ).
\eea
\ep

\medskip

The convergence of $\frac{1}{\sqrt n} \sum_{j=1}^{n} (r(\tau_j(\overline \omega))-\overline{r} ) $ for any initial measure $p_0$ s.t. $p_0(\tau)=|a_{\tau}|^2$ in distribution to $X^\omega\simeq \cN(0,\Sigma)$ implies the convergence of the characteristic function and of its derivatives, which are continuous functions of the random variable $\frac{1}{\sqrt n} \sum_{j=1}^{n} (r(\tau_j(\overline \omega))-\overline{r}) $.  In particular 
\be
-{\partial_{y_j}\partial_{y_k}e^{-iy\overline{r}\sqrt n }\Phi_n^{\ffi_0}\left(\frac{y}{\sqrt n}\right)|_{y=0}}=\sum_{\tau_0\in I_\pm}|a_{\tau_0}|^2\frac{1}{n} \left(\sum_{l=1}^{n} ( r(\tau_l(\overline \omega))-\overline{r}  )_j\ \sum_{l=1}^{n} (r( \tau_l(\overline \omega))-\overline{r}  )_k\right),
\ee
whose limit, as $n\ra\infty$ yields the elements of the random diffusion matrix.

We have
\begin{cor}\label{randd} Under the assumptions of Theorem \ref{clt}, the following random variables converge in distribution as $n\ra\infty$:  \\
The random rescaled characteristic functions
\bea
e^{-iy\overline{r}\sqrt n }\Phi_n^{\ffi_0}(y/\sqrt n)&=&\sum_{\tau_0\in I_\pm}|a_{\tau_0}|^2\left(e^{iy\frac{1}{\sqrt n} \sum_{j=1}^{n} (r(\tau_j(\overline \omega)) -\overline{r} )}\right)\longrightarrow e^{iyX^\omega} ,\\
e^{-iy\overline{r}\sqrt n }\Phi_n^{\rho_0}(y/\sqrt{n})&=&\sum_{\tau_0\in I_\pm }e^{iy \frac{1}{\sqrt n}\sum_{j=1}^n(r(\tau_j(\overline\omega))-\overline{r} )}\int_{\T^d} 
\bra \tau_0| R_0(y/{\sqrt n}-v,v) \tau_0\ket
\, d \tilde v \nonumber\\
&&\longrightarrow e^{iy X^\omega}, \ \mbox{where 
$X^\omega\simeq \cN(0,\Sigma)$,}
\eea

and the random diffusion constants
\bea
&&\sum_{\tau_0\in I_\pm}|a_{\tau_0}|^2\frac{1}{ n} \left(\sum_{l=1}^{n} ( r(\tau_l(\overline \omega) )-\overline r)_{j}\sum_{l=1}^{n}  ( r(\tau_l(\overline \omega) )-\overline r)_{k}\right)\longrightarrow \D_{jk}^{\omega},\\
&&\sum_{\tau_0\in I_\pm }\int_{\T^d} 
\bra \tau_0| R_0(y/{\sqrt n}-v,v) \tau_0\ket
\, d \tilde v\frac{1}{ n} \left(\sum_{l=1}^{n} ( r(\tau_l(\overline \omega) )-\overline r)_{j}\sum_{l=1}^{n}  ( r(\tau_l(\overline \omega) )-\overline r)_{k}\right)
\nonumber\\
&&\quad \quad \quad\quad \quad \quad \quad\quad \quad \quad\quad \quad\quad \quad \quad\quad \quad \quad \quad \quad \ \ \longrightarrow \D_{jk}^{\omega},
\eea
where $\D_{jk}^\omega$ is distributed according to the law of $X_j^\omega X_k^\omega$, where 
$X^\omega\simeq \cN(0,\Sigma)$.
\end{cor}

\begin{rem} In particular we get
\be
\E_\omega(\D_{jk}^\omega)=\Sigma_{jk}.
\ee
\end{rem}
\proof
For the case of initial density matrix $\rho_0$, it is enough to note that
\be
\sum_{\tau_0\in I_\pm }\int_{\T^d} 
\bra \tau_0| R_0(-v,v) \tau_0\ket
\, d \tilde v =\int_{\T^d} \tr R_0(-v,v)\, d \tilde v =\Phi_0(0)=1.
\ee
One concludes using the convergence results stated  in \cite{bil}, p. 28 and 30.\ep

\medskip

At this point one may wonder if the assumption $P=\E_\mu(C^T(\omega))$ is enough to apply Theorem \ref{cf} and compare the results concerning the averaged distribution $w(n)=\E(W^\omega(n))$. The next proposition answers this question positively
\begin{prop}
Under the hypotheses of Theorem \ref{clt}, assumption S' holds. Moreover,  the diffusion constant $\D(v)$ given by Theorem \ref{cf} is independent of $v\in \T^d$.
\end{prop}
\proof
We need to consider 
\be
\cM(y,y')=D(y,y')\cE=\sum_{\tau, \tau'\in I_\pm}e^{i(yr(\tau)+y'r(\tau'))}|\tau\otimes \tau'\ket\bra \tau\otimes \tau' | \ \cE,
\ee
 where $\cE=\E_\nu(C_\omega \otimes \overline{C_\omega)}$, with $\nu$ and $C(\omega)$ defined above (\ref{mmm}).
We first observe that the $\cM^*(Y)$-cyclic subspace generated by ${\bf \Psi}_1$, $\cI$, is given by
$
\cI=\mbox{span} \{|\sigma\otimes\sigma \ket\}_{\sigma\in I_\pm}.
$
Indeed, ${\bf \Psi}_1\in \cI$ and $\cI$ is invariant under $D(Y)$ for any $Y=(y,y')\in \T^{2d}$ since $D(y,y')|\sigma\otimes\sigma \ket=e^{i(y+y')r(\sigma)}|\sigma\otimes\sigma \ket$. Then, for any 
 $C(\pi, \Theta)=\sum_{\tau\in I_\pm}e^{i\theta_{\pi(\tau)}}|\pi(\tau)\ket\bra \tau |$ and any $\{\alpha_\sigma\}_{\sigma\in I_\pm}$, $\alpha_\sigma\in \C$, we compute
 \bea\label{samesame}
 C(\pi, \Theta)\otimes \overline{C(\pi, \Theta)} \ \sum_{\sigma\in I_\pm}\alpha_\sigma |\sigma\otimes\sigma \ket&=&\sum_{\sigma\in I_\pm}\alpha_\sigma  |\pi(\sigma)\otimes\pi(\sigma) \ket \nonumber\\
 &\equiv& C(\pi)\otimes \overline{C(\pi)} \ \sum_{\sigma\in I_\pm}\alpha_\sigma |\sigma\otimes\sigma \ket.
 \eea
This shows that $\cI$ is invariant under $(C_\omega \otimes \overline{C_\omega})^*$, and thus under its expectation as well,  which is enough to prove the claim. Moreover, (\ref{samesame}) shows that, when restricted to $\cI$, any matrix $C(\pi, \Theta)\otimes \overline{C(\pi, \Theta)} $ acts like $C(\pi)\otimes \overline{C(\pi)}$ does. Consequently, we can consider the finite measure $\mu$ on ${\mathfrak S}_{2d}$ defined by (\ref{mmm}) instead of the original measure $\nu$. Altogether we get
\bea
\cM(-v,v)|_{\cI}&=&D(-v,v)|_{\cI}\sum_{\pi\in {\mathfrak S}_{2d}}\mu(\pi)C(\pi)\otimes \overline{C(\pi)}|_{\cI}
\nonumber\\
&=&\sum_{\pi\in {\mathfrak S}_{2d}}\mu(\pi)C(\pi)\otimes {C(\pi)}|_{\cI}=\cE|_{\cI},
\eea
which is independent of $v\in \T^d$. It remains to show that ${\bf \Psi}_1$ is the only invariant vector under $\cE|_{\cI}$. But the equation for the coefficients $\alpha_\sigma$
\be
\sum_{\pi\in {\mathfrak S}_{2d}}\mu(\pi)C(\pi)\otimes {C(\pi)}\sum_{\sigma\in I_\pm}\alpha_\sigma |\sigma\otimes\sigma \ket=\sum_{\sigma\in I_\pm}\alpha_\sigma |\sigma\otimes\sigma \ket
\ee
is equivalent to $\sum_{\pi\in {\mathfrak S}_{2d}}\mu(\pi)\alpha_{\pi^{-1}(\sigma)}=\alpha_\sigma$, for all $\sigma$. This is in turn equivalent to requiring that the vector $\alpha \in \C^{2d}$ with components $\{\alpha_\sigma\}_{\sigma\in I_\pm}$ be invariant under $P^T$. $P$ being irreducible by assumption, the only invariant vectors have constant components and thus are proportional to ${\bf \Psi}_1$.
\ep
\begin{rem}
The assumption S doesn't always hold for the case under study. In case $\Theta\equiv 0$, the vector ${\bf \Psi}=\sum_{(\tau, \tau')\in I^2_\pm}|\tau\otimes\tau'\ket$ is also invariant under $\cM(0,0)=\cE$.
\end{rem}

\bigskip

We close this section by providing a general relationship between the diffusion matrix $\D^\omega$ computed by means of $W^\omega(n)$ and the diffusion matrix $\D$ computed by means of the averaged distribution $w(n)=\E_\omega(W^\omega(n))$ in Theorem \ref{cf}, provided both exist.
\medskip

To deal with the whole diffusion matrix at once, we let $z\in\R^d\setminus\{0\}$ be fixed and consider the non-negative random variables  
\be\label{rdl}
D=\bra z | \D z\ket,   \ \ D_n^\omega=\bra z |Y_n^\omega\ket\bra Y_n^\omega | z\ket , \ \ \mbox{where} \ \ Y_n^\omega =\frac{X_n^\omega-n\overline r}{\sqrt n}.
\ee
With these notations, $\E_{w(n)}(D_n)=\E_\omega\E_{W^\omega(n)}(D_n^\omega)\ra D$ by Theorem \ref{cf} and $\E_{W^\omega(n)}(D_n^\omega)\ra  D^\omega$ where $D^\omega=\bra z|\D^\omega z\ket$  as $n\ra \infty$, if the limit exists.
\begin{prop} Consider the initial condition $\Psi_0=\ffi_0\otimes |0\ket$ and
assume the hypotheses of Theorem \ref{cf} hold for the distribution $w(n)$. Further assume the random variables defined in (\ref{rdl}) satisfy $\E_{W^\omega(n)}(D_n^\omega)\ra D^\omega$ in distribution. Then
\be
\E_\omega(D^\omega)=\lim_{n\ra\infty}\E_\omega\E_{W^\omega(n)}(D_n^\omega),
\ee
which implies $\E_\omega(\D^\omega)=\D$.
\end{prop}
\proof
Let $d_n^\omega=\E_{W^\omega(n)}(D_n^\omega)\geq 0$ which converges in distribution to $D^\omega$. By Theorem 5.4 of \cite{bil}, if $\sup_{n}\E_\omega((d_n^\omega)^{1+\epsilon})<\infty$, for some $\epsilon>0$, then the limit and expectation commute:  $\lim_{n\ra\infty}\E_\omega(d_n^\omega)=\E_\omega(D^\omega)$, which yields the result. Now, Remark \ref{more} below Theorem \ref{cf} implies the condition with $\epsilon=1$.
\ep

\begin{rem} When applied to the case discussed in the section, the proposition yields
\be
\E_\omega(\D^\omega)=\Sigma=\int_{\T^d}\D\, d\tilde v =\D.
\ee
\end{rem}

\subsection{Large Deviations}

We can complete the picture for the model at hand by looking at its large deviations properties. Because the analysis reduces to the study of a finite state Markov chain, a large deviation principle is true for the model, see \cite{dz}, Theorem 3.1.2. 
\medskip 

For $\lambda\in \R^d$, let $\Pi_\lambda\in M_{2d}(\R^+)$ be the matrix whose non-negative elements  read
\be
\Pi_\lambda(\sigma,\tau)=P(\sigma, \tau)e^{\bra\lambda | r(\tau)\ket},
\ee
where $P$ is the transition matrix and $r$ is the jump function. As $P$ is irreducible, $\Pi_\lambda$ is irreducible as well. Hence, by Perron-Frobenius theorem, for all $\lambda\in\R^d$, the largest real eigenvalue of $\Pi_\lambda$, $\rho(\lambda)>0$, is simple and $\sigma(\Pi_\lambda)\subset D(0, \rho(\lambda))$. As a function of $\lambda\in \R^d$, $\rho(\lambda)$ is real analytic. For every $x\in \R^d$, define
\be
I(x)=\sup_{\lambda\in \R^d}(\bra \lambda | x\ket -\ln(\rho(\lambda))).
\ee
The function $I$ is a {good rate function}. \begin{thm}\label{ldp}
Under the hypotheses of Theorem \ref{clt}, the random variable $Z_n=\frac1n\sum_{j=1}^{n}r(\tau_j)$ satisfies a large deviation principle with convex good rate function $I$: For any $\sigma\in I_\pm$ and any $\Gamma \subset \R^d$
\bea
-\inf_{x\in \Gamma^0}I(x)&\leq &\liminf_{n\ra\infty}\frac{1}{n}\ln (\P_\sigma(Z_n\in \, \Gamma))\nonumber\\
&\leq&\limsup \frac{1}{n}\ln (\P_\sigma(Z_n\in \Gamma))\leq -\inf_{x\in \overline{\Gamma}}I(x),
\eea
where $\P_\sigma$ refers to the initial law $p_0(\tau)=\delta_{\sigma, \tau}$ 
\end{thm}

\subsection{Example}

We end this section by a simple example which allows us to make explicit all quantities encountered so far.

\bigskip

In dimension $d=1$, we consider the jump function $r$ defined by $r(\pm 1)=\pm 1$, so that $\overline r=0$. We define the distribution $\mu$ on ${\mathfrak S}_2$ by the Bernoulli process which assigns probability $p>0$ to the identity matrix and $q=1-p$ to the matrix $\begin{pmatrix}0 & 1 \\ 1 & 0 \end{pmatrix}$. The corresponding transition matrix reads 
\be
P=\begin{pmatrix}p & q \\ q & p \end{pmatrix} \ \mbox{s.t. } \ (P-z)^{-1}=\frac{|{\bf \Psi}_0\ket\bra{\bf \Psi}_0|}{1-z} + \frac{|\psi_1\ket\bra\psi_1|}{(p-q)-z}, 
\ee
with ${\bf \Psi}_0=\frac{1}{\sqrt 2}\begin{pmatrix}1 \\ 1\end{pmatrix}$, $\psi_1=\frac{1}{\sqrt 2}\begin{pmatrix}1 \\ -1\end{pmatrix}$. Consequently $S(1)=\frac{|\psi_1\ket\bra\psi_1|}{(p-q)-1}$. Thus, the averaged diffusion constant $\D=\Sigma$ computed from (\ref{covma}) with $r=\begin{pmatrix}1 \\ -1\end{pmatrix}$ reads
\be
\Sigma=-\frac12 \bra r|r\ket-\frac12 2\bra r|S(1)r\ket=\frac pq.
\ee
Therefore the random variable $\E_{W^\omega(n)}(\frac{X_n^\omega}{\sqrt n})$ converges in distribution to $X^\omega \simeq \cN(0,p/q)$ and the corresponding random diffusion constant $\D^\omega$  is distributed according to the chi-square law  with density $\frac pq f(\cdot \frac pq)$ where
\be
f(t)=\frac{e^{-t/2}}{\sqrt{2\pi t}}, \ \ t\geq 0.
\ee
Next we compute the rate function $I$ such that $\P_\sigma(\E_{W^\omega(n)}(\frac{X_n^\omega}{ n})\in \Gamma)\simeq e^{-n \inf_{x\in \Gamma}I(x)}$, for $n$ large. With our choice of jump function, the matrix $\Pi_\lambda$ reads
\be
\Pi_\lambda=\begin{pmatrix}pe^\lambda & qe^{-\lambda} \\ qe^{\lambda} & pe^{-\lambda} \end{pmatrix}  \ \mbox{s.t.} \ \det \Pi_\lambda = p-q \  \ \mbox{and } \ \ \tr \ \Pi_\lambda = 2 p \cosh(\lambda).
\ee
Thus we have
\be
\rho(\lambda)=p\cosh(\lambda)+\sqrt{p^2\sinh^2(\lambda)+q^2}
\ee
so that $\sup_{\lambda\in \R}(x\lambda-\ln(\rho(\lambda)))$ is reached at $\lambda(x)=\mbox{arsinh} \left(\frac{qx}{p\sqrt{1-x^2}}\right)$ for $|x|<1$. Hence
\be
I(x)=x\ \mbox{arsinh} \left(\frac{qx}{p\sqrt{1-x^2}}\right)-\ln\left(\frac{q+\sqrt{p^2+x^2(q-p)}}{\sqrt{1-x^2}}\right)  \ \mbox{if $|x|<1$ } 
\ee
and $I(x)=\infty$ otherwise.

\section{Generalization}\label{mark}

In the light of the last example, it is natural to generalize the results of Section \ref{rands} to the case where the random coin matrices $C(\omega)$ are distributed according to a Markov process, in the spirit of \cite{p}, \cite{ks}, \cite{hks}. We briefly do so in this last section, mentioning the main modifications only and considering finitely many coin matrices for simplicity. 

\medskip

Consider a finite set $\{C_1, C_2, \cdots, C_F\}$ of unitary coin matrices on $\C^{2d}$ and assume that for any $n\in \N$, the  set of random matrices $\{C(\omega_1), \cdots, C(\omega_n)\}$ is determined by a Markov chain characterized by an initial distribution $\{p_0(j)\}_{j=1}^F$ and an irreducible transition matrix $\{P(j,k)\}_{j,k\in\{1,\cdots, F\}}$. Correspondingly, for any $Y\in\T^{d}\times \T^d$, the sequence of matrices $\{M_{\omega_j}(Y)\}_{j\in\N}$ has the same distribution, assuming the $C_j\otimes C_k$ are distinct, so that 
\bea
&&\P(\{M_{\omega_n}(Y),M_{\omega_{n-1}}(Y),\cdots M_{\omega_1}(Y)\}=\{M_{k_n}(Y),M_{k_{n-1}}(Y),\cdots M_{k_1}(Y)\})\nonumber\\
&&\hspace{7cm}=p_0(k_1)P(k_1,k_2)\cdots P(k_{n-1}, k_n).
\eea
We introduce the matrices 
\be
\M_{jk}(Y)=P^T(j,k)M_k(Y), \ \ \ j,k \in\{1,\cdots, F\},
\ee acting on $\C^{2d}\otimes\C^{2d}$ and the operator acting on $\C^F\otimes (\C^{2d}\otimes\C^{2d})$
\be
\M(Y)=\sum_{j,k \in\{1,\cdots, F\}} |e_j\ket\bra e_k|\otimes \M_{jk}(Y),
\ee
where the $e_j$'s are the canonical basis vectors of $\C^F$ and the "bra"s and "ket"s refer to the usual scalar product in $\C^F$. Similarly, let $\M_0(Y)$ be the vector in $M_{4d^2}(\C)^F$ defined by
\be
\M_0(Y)=\sum_{j \in\{1,\cdots, F\}}p_0(j)|e_j\ket\otimes M_j(Y).
\ee
If $\chi_1=\sum_{j \in\{1,\cdots, F\}}|e_j\ket \in \C^F$ and $\I$ is the identity operator on $\C^{2d}\otimes\C^{2d}$, we obtain 
\be
\E(M_{\omega_n}(Y)M_{\omega_{n-1}}(Y)\cdots M_{\omega_1}(Y))=\bra \chi_1\otimes \I | \M(Y)^{n-1}\M_0(Y)\ket.
\ee
Hence, with ${\bf \Psi}_1=\sum_{\tau\in I_\pm}|\tau\otimes\tau\ket$ and $\Phi_0\in \C^{2d}\otimes\C^{2d}$, we can write
\be
\E(\Phi_n^{\overline\omega}(Y))=\bra \chi_1\otimes {\bf \Psi}_1 | \M(Y)^{n-1}\ \M_0(Y) \Phi_0\ket,
\ee
where $\chi_1\otimes {\bf \Psi}_1$ and $\M_0(Y) \Phi_0=\sum_{j }p_0(j) |e_j\ket \otimes M_j(Y)\Phi_0$ belong to $\C^F\otimes(\C^{2d}\otimes\C^{2d})$ and "bra"s and "ket"s should be interpreted accordingly. 

\medskip

This brings us back to the study of large powers of an operator, $\M(Y)$, which has essentially the same properties as $\cM(Y)$:

\begin{lem} The matrix $\M(Y)$ acting on $\C^F\otimes(\C^{2d}\otimes\C^{2d})$ is analytic in $Y\in \C^{2d}\times \C^{2d}$.\\
Assume $P$ is irreducible, and let $\chi_p\in \C^F$ be the unique real valued vector s.t. $P^T\chi_p=\chi_p$ and $\bra \chi_p|\chi_1\ket=1$. Then for any $v\in \T^d$, 
\bea
\M(-v,v) | \chi_p\otimes {\bf \Psi}_1\ket= |\chi_p\otimes {\bf \Psi}_1\ket \ \ \mbox{and} \ \
\M^*(-v,v) | \chi_1\otimes {\bf \Psi}_1\ket= |\chi_1\otimes {\bf \Psi}_1\ket.
\eea
For any $Y\in \T^d\times\T^d$, 
\be
{\em \spr} \M(Y)={\em \spr} \M^*(Y)\leq 1.
\ee
\end{lem}
\proof
The first two identities follow from explicit computations. The second property is a consequence of the fact that if one endows $\C^F\otimes(\C^{2d}\otimes\C^{2d})$ with the norm $\| \sum_{j}e_j\otimes\Psi_j\|_\infty:=\max_{j}\|\Psi_j\|_{\C^{2d}\otimes\C^{2d}}$, then $\M^*(Y)$ becomes a contraction. This is due to the fact that $P$, as a stochastic matrix, is a contraction with the sup norm on $\C^F$.  The spectral radius of $\M^*(Y)$ thus cannot exceed one.
\ep\\

We shall work under the \\

{\bf Assumption S'':} For all $v\in \T^d$, 
\be
\sigma(\M(-v,v)|_{\cI^*})\cap \partial D(0,1)=\{ 1\} \ \ \mbox{and the eigenvalue $1$ is simple,}  
\ee
where $\cI^*$ is the $\M^*(Y)$-cyclic subspace generated by $\chi_1\otimes {\bf \Psi}_1$.\\

Then we can proceed with a spectral analysis similar to that of Section \ref{rands}. Ignoring the restriction $|_{\cI^*}$ in the notation, we note that Assumption S'' implies that for all $v \in \cT_{\nu}^d$, a complex neighborhood of $\T^d$, we can write 
\be
(\M(-v,v)-z)^{-1}=\frac{\tilde P}{1-z}+\tilde S_v(z), \ \ z\not\in \sigma(\M(-v,v)),
\ee
where $\tilde P = \frac 1d | \chi_p\otimes {\bf \Psi}_1\ket\bra  \chi_1\otimes {\bf \Psi}_1|$ is independent of $v$, and the reduced resolvent $\tilde S_v(z)$ has the same analyticity properties as that of $\cM(-v,v)$. Moreover, introducing 
\be
\Delta(Y)=\sum_j |e_j\ket\bra e_j|\otimes D(Y)\ee
we see that $\M(y-v,v)=\Delta(y,0)\M(-v,v)$ where
\be
\Delta(y,0)=\sum_j |e_j\ket\bra e_j|\otimes (\I + F_1(y)+F_2(y)+O(\|y\|^3)),
\ee
see (\ref{fy3}). Therefore, if $(y,v)\in \cB(0,y_0)\times \cT_{\nu}^d$, for $y_0>0$ and $\nu>0$ small enough, Lemma \ref{specprop} holds for $\M(y-v,v)$. Applying the same perturbation formulas for the isolated eigenvalue of $\M(y-v,v)$,  noted $\lambda_1(y,v)$ again, for $y\in \C^d$ small enough, we reach the same conclusions by explicit computations:
\bea\label{difmatmark}
\lambda_1(y,v)&=&1+\frac{i}{2d}\sum_{\tau\in I_\pm}yr(\tau)-\frac 12 \bra y | \D(v) y\ket+ O_v(\|y\|^3),
\eea
for all $v \in {\cT}_{\nu}^{d}$.
The first order term in $y$ is the same as the one of (\ref{difmat}). On the other hand, the explicit form of the quadratic term in $y$ which defines the (analytic) diffusion matrix $\D(v)$, depends on the transition matrix $P$
\bea
\bra y |\D(v) y\ket&=&-\frac1d \sum_{\tau\in I_\pm}\frac{(yr(\tau))^2}{2}
\\ \nonumber
& &-\frac{1}{d}\left(  \sum_{\tau, \tau'\in I_\pm} (yr(\tau))(yr(\tau'))\left\{
\bra \chi_1\otimes \tau\otimes\tau |\tilde S_v(1) \chi_p\otimes\tau'\otimes \tau'\ket
-\frac{1}{2d}\right\} \right).
\eea
 Moreover, the corresponding rank one projector $\tilde P(y,v)$ is analytic and thus tends to $\tilde P$ uniformly in $v\in {\cT}_{\nu}^{d}$, as $y\ra 0$. With  $\tilde Q(y,v)=\I-\tilde P(y,v)$, the matrix $\tilde Q(y,v)\M(y-v,v)\tilde Q(y,v)$ has spectral radius strictly small than one, for $v\in {\cT}_{\nu}^{d}$ and $y\in \C^{d}$ small enough. 

\medskip
Therefore we can state that all conclusions drawn in Section \ref{rands}, e.g. Theorem \ref{cf}, and in Section \ref{moder}, e.g. Theorem \ref{md}, for i.i.d coin matrices under Assumption S are true for finitely many coin matrices forming a Markov chain, under Assumption S'', {\it mutatis, mutandis}.

\end{document}